%% file: igseek.tex
\documentclass{article}
\usepackage{GEM_workshop_2025,times}

\input{math_commands.tex}

\usepackage{hyperref}
\usepackage{url}
\usepackage{algorithm}
\usepackage{algorithmic}
\usepackage{epstopdf}
\usepackage{graphicx}
\usepackage{subfigure}
\usepackage{booktabs}
\usepackage{float}            
\usepackage{overpic}
\usepackage{comment}
\usepackage{multirow}
\usepackage{amsthm}
\usepackage{wrapfig}
\usepackage{xcolor}
\usepackage{array}
\usepackage{picinpar}
\usepackage{footmisc}
\usepackage{subcaption}
\usepackage{enumitem}
\usepackage{multirow}
\usepackage{bbding}
\usepackage{colortbl}
\usepackage{wrapfig} 
\usepackage[textsize=small]{todonotes}
\definecolor{LightSteelBlue}{RGB}{213,229,255}
\input{symbol}

\title{Fast and Accurate Antibody Sequence Design via Structure Retrieval}

\author{
  Xingyi Zhang$^{1}$\thanks{Equal contribution} \quad Kun Xie$^{2,3}$\samethanks[1] \quad Ningqiao Huang$^{2}$ \quad Wei Liu$^{2}$ \quad Peilin Zhao$^{2}$ \\ \hspace{0.1mm} \textbf{Sibo Wang}$^{3}$ \quad \textbf{Kangfei Zhao}$^{4}$\thanks{Corresponding authors.} \quad \textbf{Biaobin Jiang}$^{2}$\samethanks[2]  \\
  $^{1}$MBZUAI \quad $^{2}$Tencent AI Lab \quad $^{3}$The Chinese University of Hong Kong \\ $^{4}$Beijing Institute of Technology\\
  \hspace{0.1mm} \texttt{zkf1105@gmail.com} \quad
  \texttt{brunojiang@tencent.com}
}

\iclrfinalcopy
\begin{document}

\maketitle

\input{section/0-abstract}

\input{section/1-introduction}

\input{section/5-result}

\input{section/6-conclusion}

\bibliography{igseek}
\bibliographystyle{iclr2025_conference}

\input{section/7-appendix}

\end{document}

%% file: math_commands.tex
\usepackage{amsmath,amsfonts,bm}

\def\eqref#1{equation~\ref{#1}}

\def\1{\bm{1}}

\DeclareMathAlphabet{\mathsfit}{\encodingdefault}{\sfdefault}{m}{sl}
\SetMathAlphabet{\mathsfit}{bold}{\encodingdefault}{\sfdefault}{bx}{n}



%% file: symbol.tex
\newcommand*\samethanks[1][\value{footnote}]{\footnotemark[#1]}

\newcommand{\vect}[1]{\boldsymbol{#1}}
\def\igseek{IgSeek}

\newtheorem{theorem} {Theorem}

%% file: section/0-abstract.tex
\begin{abstract}
\label{sec:abstract}
Recent advancements in protein design have leveraged diffusion models to generate structural scaffolds, followed by a process known as protein inverse folding, which involves sequence inference on these scaffolds. However, these methodologies face significant challenges when applied to hyper-variable structures such as antibody Complementarity-Determining Regions (CDRs), where sequence inference frequently results in non-functional sequences due to hallucinations. Distinguished from prevailing protein inverse folding approaches, this paper introduces {\igseek}, a novel structure-retrieval framework that infers CDR sequences by retrieving similar structures from a natural antibody database. Specifically, {\igseek} employs a simple yet effective multi-channel equivariant graph neural network to generate high-quality geometric representations of CDR backbone structures. Subsequently, it aligns sequences of structurally similar CDRs and utilizes structurally conserved sequence motifs to enhance inference accuracy. Our experiments demonstrate that {\igseek} not only proves to be highly efficient in structural retrieval but also outperforms state-of-the-art approaches in sequence recovery for both antibodies and T-Cell Receptors, offering a new retrieval-based perspective for therapeutic protein design.
\end{abstract}

%% file: section/1-introduction.tex
\section{Main}
\label{sec:intro}

Antibodies, known for their high specificity and affinity, have emerged as pivotal therapeutic agents in the treatment of complex diseases, including cancer \cite{adams2005cancer}, autoimmune disorders \cite{feldmann2003tnf}, and infectious diseases \cite{abraham2020passive}. In 2023, the global best-selling drug was Keytruda, a cancer treatment antibody, with sales reaching $\$25$ billion, surpassing Humira, another antibody used for treating rheumatoid arthritis, which had dominated the market for the past decade \citep{dunleavy2024keytruda}. Traditionally, the discovery of antibodies has predominantly relied on immunizing animals with antigens \cite{van1980okt3} or employing various display techniques such as phage \cite{maccallum1996antibody} and yeast displays \cite{chao2006yeast}. However, these approaches face significant challenges when dealing with structurally intricate proteins, which are difficult to express in a soluble and functional form. Additionally, even when numerous candidate antibodies are generated through these techniques, they may not necessarily bind to the desired domain or exhibit therapeutic efficacy.

To overcome these limitations, deep learning models have been introduced to design synthetic antibodies by learning from natural antibody-antigen complexes \cite{luo2022diffab,jin2022refinegnn,kong2023dymean,kong2023mean,bennett2024atomically}. Despite significant strides in protein design \cite{dauparas2022mpnn,hsu2022esmif1,notin2024protein}, antibodies present a distinct challenge for deep learning due to the high flexibility of their binding regions, known as complementarity-determining regions (CDRs). Inspired by RFdiffusion's \cite{watson2023rfdiffusion} remarkable achievements in monomeric protein and binder design, Bennett et al. \cite{bennett2024atomically} advanced the field by fine-tuning the RFdiffusion model with antibody-antigen complex structural data to facilitate epitope-targeted antibody design. 
Their approach aligns well with established pharmaceutical practices by generating different CDRs on the same framework for different antigen targets, thereby enhancing developability and reducing downstream optimization requirements. 
While structural and functional analyses validated the its capability to generate antibodies that bind to predetermined epitopes, the approach was constrained by notably low success rates.

One reason for the low success rate of this AI-based antibody design pipeline is the occurrence of hallucinations, 
especially during the process of protein inverse folding, which predicts the CDR sequence based on the backbone structure \cite{dauparas2022mpnn,hsu2022esmif1,gruver2023lambo,gao2023pifold}. To be specific, given an antigen epitope and an antibody backbone,
the amino acid sequences inferred through methods like ProteinMPNN \cite{dauparas2022mpnn} and ESM-IF1 \cite{hsu2022esmif1} may not fold into the desired structures in real biological systems. More critically, there are currently no effective computational methods to reduce these hallucinations, aside from conducting time-consuming, labor-intensive, and expensive wet-lab experiments for validation. Typically, using independent structure prediction models to fold and verify the inferred sequences cannot effectively eliminate non-functional sequences caused by hallucinations. That is because even state-of-the-art models exhibit structural deviations of 1 to 3 \AA\ and have low confidence in predicting the structures of antibody CDRs. 

\begin{wrapfigure}{r}{0.5\textwidth}
  \centering
  \hspace{-2mm}
  \includegraphics[width=\linewidth]{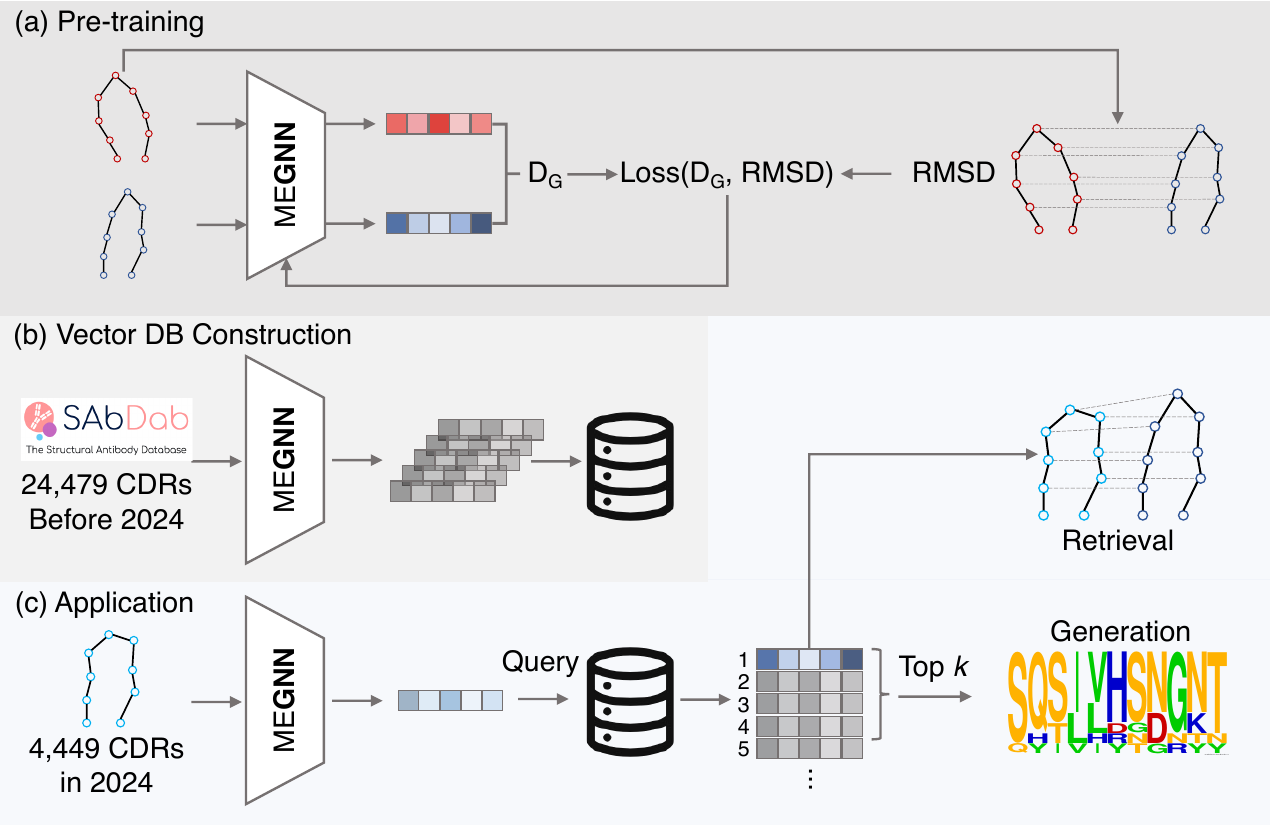}
  \vspace{-3mm}
  \caption{The Framework IgSeek: (a) Pre-train an MEGNN encoder by a self-supervised learning task. (b) Construct a CDR vector database. (c) Sequence generation by K-NN search.}
  \label{fig:framework}
  \vspace{-4mm}
\end{wrapfigure}
To deal with the challenge of hallucinations arising from previous models, we propose an antibody CDR sequence design framework from a novel perspective of similar structure retrieval, named as  {\igseek} (Ig for Immunoglobulin, a.k.a. antibody). Our framework is enlightened by a noteworthy empirical discovery made 25 years ago, which revealed that antibodies exhibit a limited set of canonical structures within 5 out of 6 CDRs despite the vast diversity in sequences, and that certain CDR conformations are scaffolded by a few highly conserved residues \cite{chothia1989conformations}. Further inspired by retrieval-augmented prediction for hallucination reduction in protein structure prediction~\cite{jumper2021alphafold}, and natural language generation~\cite{DBLP:journals/corr/abs-2312-10997}, 
{\igseek} leverages neural retrieval in a database of natural antibodies to retrieve structurally similar sequence templates of CDR, and ensembles the queried templates for sequence prediction. 
Extensive experimental validation demonstrates that our structure-guided retrieval approach effectively improves the accuracy of CDR sequence prediction, notably outperforming state-of-the-art sequence design methods.

%% file: section/5-result.tex
\section{Results} 
\label{sec:res}

\subsection{IgSeek Approach}
\label{sec:subsec-res-setting}
The gist of IgSeek for structure-to-sequence generation is isomorphic structure retrieval, which allows for the exploration of a large and diverse antibody CDR structure database. Fig.~\ref{fig:framework} illustrates the framework of IgSeek. Given an antibody CDR database where both structures and sequences are available, IgSeek first constructs a CDR vector database, where vector embeddings index the structural proximity of the CDRs. In this offline stage, we pre-train a {\em Multi-channel Equivariant Graph Neural Network (MEGNN)} to encode the structure of CDR loops into fixed-length vectors within the CDR database. Specifically, MEGNN aligns the spatial structure distance between pairs of CDRs with equal lengths and similar conformations. Subsequently, for a CDR structure $G$ whose sequence is to be predicted, we first deploy the pre-trained MEGNN to generate an embedding $\vect{h}_G$ for $G$. $\vect{h}_G$ then serves as the search key to query the {\em $K$-nearest neighbors ($K$-NN)} structurally similar CDR loops in the vector database. Finally, the $K$-NN results, associated with their corresponding residue sequences, are collected to predict the sequence of $G$ by ensemble and Bernoulli sampling.A detailed description of our methodology can be found in Appendix~\ref{sec:method}.

{\bf Datasets.} 
We evaluate our {\igseek} and other baselines using both solved and predicted antibody structures. The training set consists of CDR pairs sampled from $11,023$ solved CDR loops in the {\em Structural Antibody Database (SAbDab)} \citep{dunbar2013sabdab,schneider2021sabdab}. To construct the CDR vector database, we utilize $24,479$ solved CDR loops from SAbDab before January 1, 2024 (SAbDab-before-2024). In addition, $4,449$ solved CDR loops released between January 3, 2024 and May 29, 2024 from SAbDab (SAbDab-2024) serve as the test set to evaluate the performance of {\igseek} and its competitors.
In addition to the solved antibody structures from SAbDab, we also conduct experiments on $5,111$ CDR loops from the {\em Structural T-Cell Receptor Database (STCRDab)} \citep{leem2018stcrdab} to evaluate the model generalization ability. Furthermore, we evaluate the model efficiency using $5,000$ predicted CDR-H3 loops from the {\em Observed Antibody Space (OAS-H3)} \citep{kovaltsuk2018oas,olsen2022oas}. IMGT numbering scheme \cite{lefranc2003imgt} is utilized for antibody datasets. More details of datasets and experiment settings can be found in Appendix \ref{sec:appendix-data} and Appendix~\ref{sec:appendix-param}.

\begin{figure*}[t]
\centering
    \includegraphics[width=0.95\columnwidth]{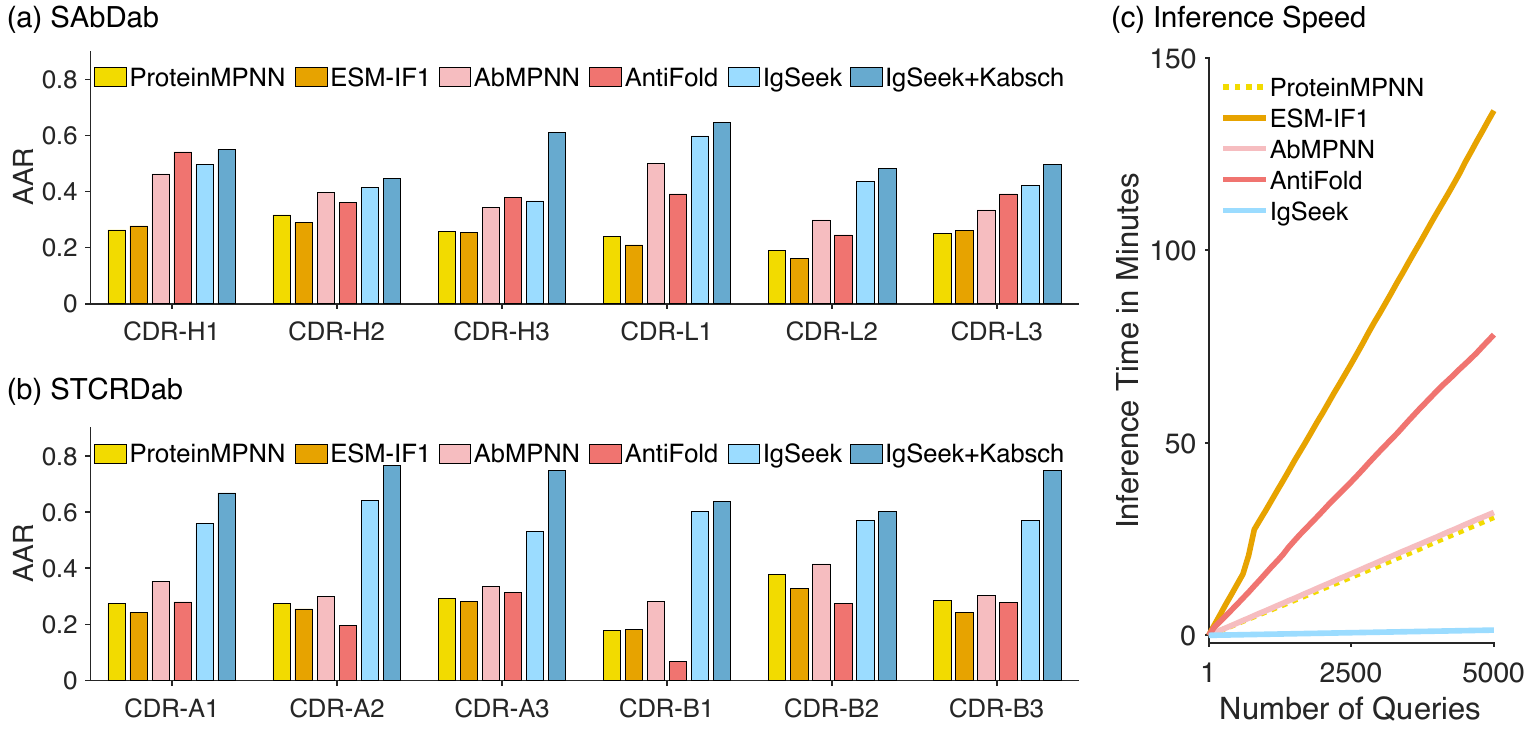}
\vspace{-3mm}
\caption{The comparison of average AAR and inference speed. (a) AAR in SAbDab-2024 dataset. (b) AAR in STCRDab. (c) Inference speed. }
\vspace{-5mm}
\label{fig:exp-seq-design}
\end{figure*}

\subsection{IgSeek for CDR Structure Retrieval}
\label{sec:subsec-res-search}

\begin{wrapfigure}{r}{0.6\textwidth}
  \centering
  \vspace{-5mm}
  \hspace{-2mm}
  \includegraphics[width=\linewidth]{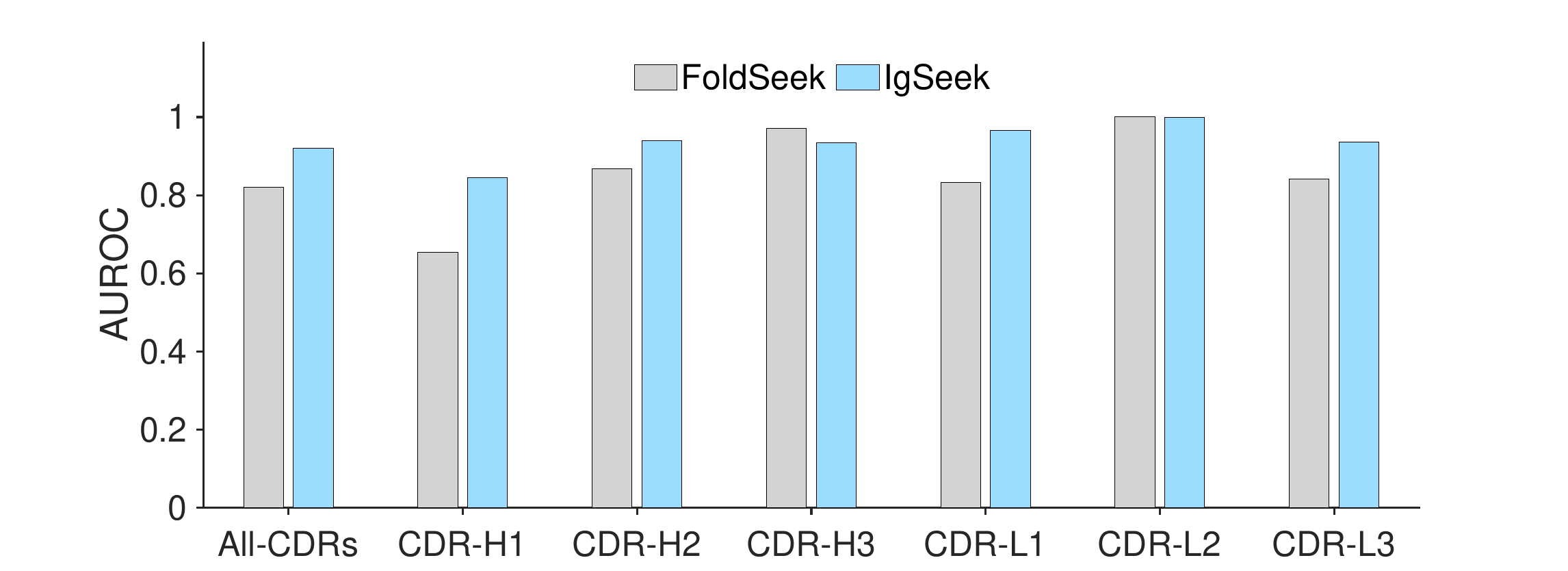}
  \vspace{-3mm}
  \caption{IgSeek vs. FoldSeek in CDR retrieval.}
  \label{fig:exp-search}
  \vspace{-3mm}
\end{wrapfigure}

In this set of experiments, we compare {\igseek} with the state-of-the-art structure searching model, FoldSeek~\cite{van2024foldseek}, by examining the quality of the retrieved isomorphic structures. Introduction to competitors is deferred to Appendix~\ref{sec:appendix-baseline}. Specifically, for a given query CDR $q$, the retrieved CDR $r$ is considered a positive instance if their RMSD is less than 1 \AA. To ensure the robustness of our evaluation, we omit any query CDR for which there are no candidates in the CDR database with a distance of less than 1 \AA\ from the query. This strategy allows us to focus on instances where meaningful comparisons can be made, thereby enhancing the result reliability.

Fig. \ref{fig:exp-search} presents the experimental results of {\igseek} and FoldSeek, illustrating the model performance on the retrieved sequences using the AUROC metric. As we can observe, {\igseek} outperforms FoldSeek on four types of CDR loops while maintaining comparable performance on CDR-H3 and CDR-L1, indicating its capability of identifying structurally similar CDRs across diverse CDR loops. It is worth noting that {\igseek} achieves a 2.6x speed-up in structure retrieval time compared to FoldSeek. Since this improvement in speed does not come at the cost of accuracy, it demonstrates that {\igseek} strikes a superior trade-off between efficiency and accuracy. The ability to quickly retrieve high-quality structural matches can greatly enhance workflows in antibody design, as shown in Sec.~\ref{sec:subsec-res-gen}.

\begin{figure*}[t]
  \centering
  \includegraphics[width=0.95\textwidth]{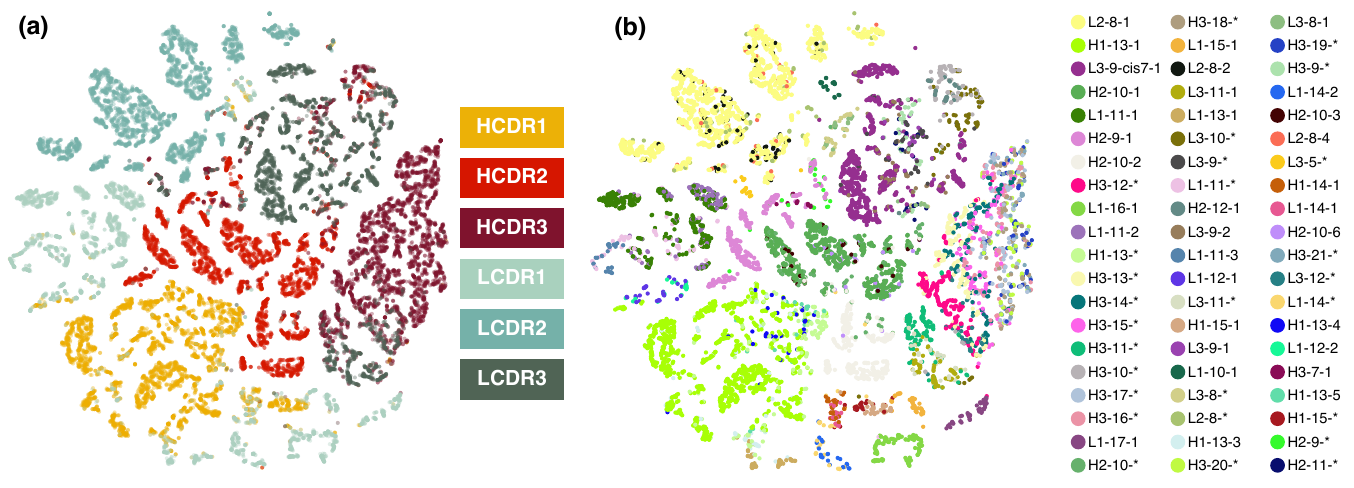}
  \vspace{-2mm}
  \caption{Embeddings of CDRs in the SAbDab-before-2024 datasets projected onto 2D Space.}
   \label{fig:exp-viz}
  \vspace{-4mm}
\end{figure*}

\subsection{IgSeek for CDR Sequence Design}
\label{sec:subsec-res-gen}
In this set of experiments, we compare {\igseek} with the state-of-the-art models for protein and antibody sequence design, including ProteinMPNN~\citep{dauparas2022mpnn}, ESM-IF1~\citep{hsu2022esmif1}, AbMPNN~\citep{dreyer2023abmpnn}, and AntiFold~\citep{hoie2024antifold}. Specifically, the MEGNN in {\igseek} is trained on the SAbDab-before-2024 dataset to construct the CDR vector database. Subsequently, the trained MEGNN is utilized to generate embeddings for the CDRs in the SAbDab-2024 dataset. For each query CDR in the SAbDab-2024 dataset, we retrieve the top-$10$ nearest neighbors from the SAbDab-before-2024 dataset in the CDR vector database, ensuring that the lengths of the retrieved sequences match that of the query. Finally, we proceed to sample the amino acids for each position in the CDR sequences to generate the predicted result for the query CDR. Note that existing protein and antibody inverse folding methodologies such as ProteinMPNN and AntiFold typically generate at least two samples for evaluation. In our experiments, we follow the settings of ProteinMPNN and present the best results of all other methods for evaluation. Average {\em amino acid recovery (AAR)} is utilized to evaluate model performance, which quantifies the accuracy of the predicted sequences. For a query CDR $q$, the AAR is defined as the ratio of overlapping positions between the predicted sequence $\hat{\vect{s}}_q$ and ground-truth sequence $\vect{s}_q$:
$
    \text{AAR}\left(\hat{\vect{s}}_q,  \vect{s}_q\right) = \frac{1}{L} \sum_{l = 1}^{L} \mathbb{I}(\hat{\vect{s}}_q(l), \vect{s}_q(l)).
$

Fig. \ref{fig:exp-seq-design} (a) illustrates the average AAR for each model on the SAbDab-2024 dataset. As we can observe, Antifold and AbMPNN achieve much better results compared to ProteinMPNN and ESM-IF1, highlighting the advantages of fine-tuning pre-trained protein design models specifically on the antibody dataset. Additionally, {\igseek} outperforms its competitors by at least 2.9\% on light chain CDR loops (CDR-L) and achieves results comparable to state-of-the-art methods on heavy chain CDR loops (CDR-H). We incorporate an additional variant of {\igseek} that uses RMSD as a secondary sorting metric, denoted as {\igseek}+Kabsch. Notice that we do not deploy the Kabsch algorithm to search the entire database. Instead, we validated the RMSD of the top-ranked CDRs identified by IgSeek until we identified the top 10 CDRs with RMSD less than 1 \AA. Notably, {\igseek}+Kabsch consistently outperforms all baselines across six types of CDR loops, highlighting the effectiveness of our retrieval-based strategy. The marked advantage of {\igseek}+Kabsch on CDR-H3 loops is particularly noteworthy, as CDR-H3 is often considered one of the most hypervariable regions.

\noindent
{\bf Remark.} We observe a performance degradation in AntiFold and AbMPNN on the SAbDab-2024 dataset compared to the results reported by \cite{hoie2024antifold}. One possible reason for this discrepancy is that these two models heavily depend on antibody backbone structures as auxiliary information, while only the structures of CDRs are given in our settings.

{\bf Generalization Performance.}
Next, we evaluate the model inference performance on the STCRDab dataset without any further model training. To conduct this evaluation, we randomly draw around $80\%$ of the CDR loops to generate selection templates, while the remaining $20\%$ are used as queries. Fig. \ref{fig:exp-seq-design} (b) displays the average AAR of each model on the STCRDab dataset. As we can see, {\igseek} takes the lead by at least $30\%$ on CDR loops from chain A and chain B, respectively. These impressive results further underscore the potential of structure retrieval approaches in mitigating hallucinations during sequence inference, demonstrating that {\igseek} can effectively generalize to unseen data while maintaining high accuracy in sequence recovery.

{\bf Efficiency Evaluation.}
We evaluate the model efficiency using the OAS-H3 dataset. Fig. \ref{fig:exp-seq-design} (c) reports the inference time of {\igseek} compared with other baseline models, all without any model retraining. As we can observe, {\igseek} achieves at least $20$x speed-up compared to baseline methods, which demonstrates that our {\igseek} achieves a better trade-off between effectiveness and efficiency. This enhanced inference speed is particularly beneficial in practical applications like high-throughput antibody design where rapid sequence generation is crucial.

{\bf Visualization. }
To investigate the representation generated by MEGNN, we conduct a visualization analysis on the SAbDab-before-2024 dataset by T-SNE \citep{van2008tsne}. Fig. \ref{fig:exp-viz} presents the visualization results of top-$60$ CDR representations in each cluster, where PyIgClassify cluster labels \citep{adolf2015fccc} (refer to Appendix \ref{sec:appendix-data}) are utilized in this set of experiments. As Fig.~\ref{fig:exp-viz} illustrates, {\igseek} produces a high-quality visualization that clearly organizes the embeddings of CDR loops from distinct clusters into separate groups with minimal overlaps. 
Furthermore, the visualization not only demonstrates the effectiveness of {\igseek} in distinguishing CDRs among different clusters but also highlights its ability to capture structural information inherent in CDR loops. This visual clarity and distinct grouping underscore the robustness and discriminative capability of {\igseek} in embedding isomorphic CDR structures closer together while ensuring distinct clusters remain well-separated, which facilitates the identification and retrieval of CDR loops based on their structural characteristics.

\begin{figure*}[t]
  \centering
  \includegraphics[width=0.95\textwidth]{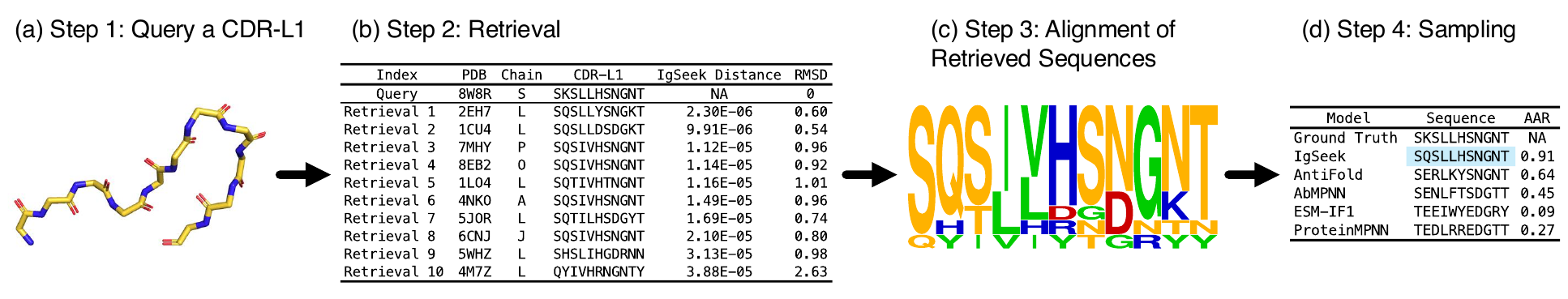}
  \vspace{-3mm}
  \caption{A Case study using 8W8R CDR-L1 as an example.}
  \label{fig:exp-case}
  \vspace{-3mm}
\end{figure*}

\input{section/tab/tab-5-dymean-comparison}

\subsection{Case Study}
\label{sec:subsec-case-study}

{\bf Example.}
We first use the 8W8R CDR-L1 as an example to illustrate the query and generation process of {\igseek}. Step 1: given the backbone structure of the 8W8R CDR-L1 loop, we employ the pre-trained MEGNN to generate its embeddings. Step 2: we retrieve the top-10 nearest neighbors of the 8W8R CDR-L1 loop from the CDR vector database $\mathcal{Z}$. Step 3: we utilize the aligned sequences from the retrieved records to generate the residue probability distribution at each position. Step 4: Finally, we sample the output result from this distribution. In this example, we observe that the AAR of the sequence generated by {\igseek} outperforms other competitors by at least 0.27, demonstrating the effectiveness of our approach.

{\bf Incorporation with structure generation models.}
Next, we evaluate the sequence prediction capabilities of Igseek using dyMEAN \cite{kong2023dymean} and RFdiffusion \cite{watson2023rfdiffusion} as structure generators for unseen antibody structures, focusing on 8R1C CDR-L1 loop. The pipeline first employs dyMEAN or RFdiffusion to generate missing CDR structures, followed by sequence predictions by Igseek for this loop, and the experimental results are presented in Table~\ref{tab:dymean-comparison}.The analysis reveals significant improvement in sequence recovery rates for 8SGN CDR-L1 when using Igseek, demonstrating the effectiveness of Igseek as a key component in real-world antibody design pipelines. In particular, our experiments indicate that the quality of structural prediction substantially influences the accuracy of sequence generation, suggesting that the development of precise CDR structure prediction models would further enhance the overall antibody design pipeline.

%% file: section/tab/tab-5-dymean-comparison.tex
\begin{table}[t]
\small
\caption{Igseek with structure generators on 8R1C CDR-L1.}
\vspace{-3mm}
\centering
\label{tab:dymean-comparison}
\begin{tabular}{lccc}
    \toprule
    Model & Sequence & AAR$\uparrow$ & RMSD$\downarrow$ \\
    \midrule
    Ground Truth & SSDVGSYNL & - & - \\ 
    dyMEAN & SSQSLLYSS & 0.33 & 4.10 \\
    dyMEAN$\Rightarrow$Igseek & SSNIGSGYD & 0.44 & 4.10\\ 
    RFdiffusion$\Rightarrow$Igseek & SSDIGAYND & {\cellcolor{LightSteelBlue}0.67} & {\cellcolor{LightSteelBlue}0.38} \\ 
    \bottomrule
\end{tabular}
\vspace{-4mm}
\end{table}

%% file: section/6-conclusion.tex
\section{Conclusion}
\label{sec:conclusion}

In this paper, we propose an antibody sequence design framework, {\igseek}, from a new learning-based structure retrieval perspective. Specifically, {\igseek} first constructs a CDR vector database using a multi-channel equivariant graph neural networks. It then predicts CDR sequences from templates retrieved from isomorphic structures in the database. Extensive experiments demonstrate the effectiveness and efficiency of {\igseek}, providing insights into de novo antibody sequence design and can inspire further investigatino in this direction.

%% file: section/7-appendix.tex
\appendix
\section*{Appendix}
\label{sec:appendix}

\input{section/2-related-work}

\input{section/3-preliminary}
\input{section/4-method}

\input{section/tab/tab-2-param}

\section{Datasets and Labels}
\label{sec:appendix-data}
{\bf Datasets.} 
We selected all experimentally solved antibody structures released in the SAbDab antibody database \citep{dunbar2013sabdab,schneider2021sabdab} before January 1, 2024, to sample our training set. Notice that we remove CDR sequences that are identical to those in the dataset to eliminate redundancy in the dataset.
Following FoldSeek \citep{van2024foldseek}, for each CDR in the SAbDab-before-2024 dataset, we randomly sample equal-length CDRs with TM-score large than $0.6$ to generate training pairs. The final training set consisted of $45,043$ antibody CDR pairs. 
After finishing model training, all $24,479$ unique CDR structures in the SAbDab-before-2024 dataset are utilized to construct the CDR vector database. 
The test set of SAbDab-2024 include experimentally solved antibody released in SAbDab antibody database between January 3, 2024 and May 29, 2024. 
This process resulted in $4,449$ test CDR samples that are completely unseen during the model training process.

\begin{figure}[h]
  \centering
  \includegraphics[width=0.55\textwidth]{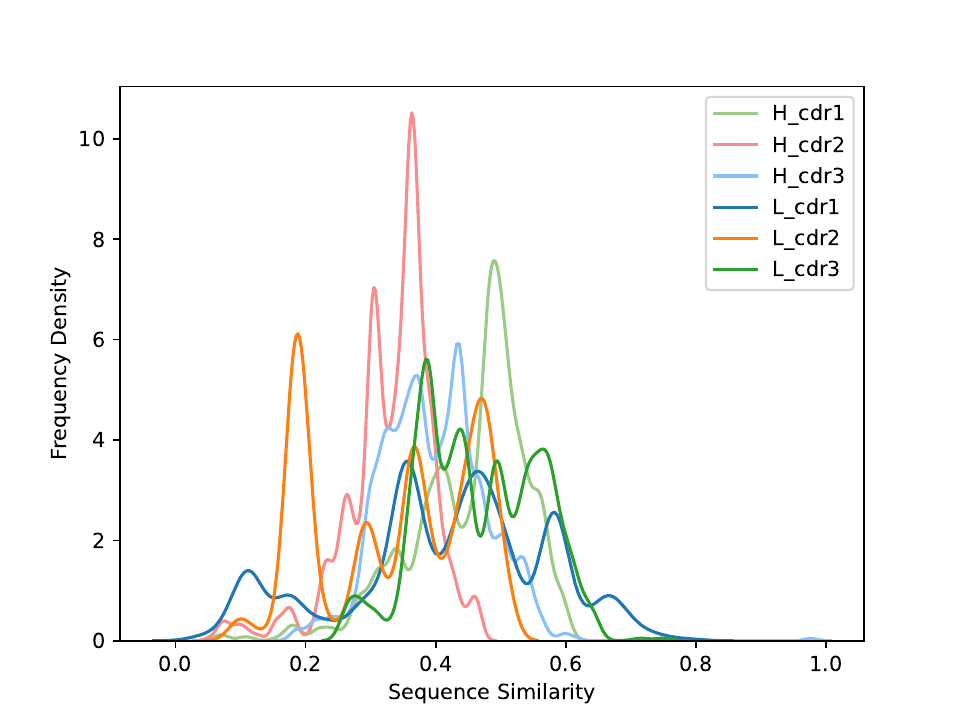}
  \vspace{-4mm}
  \caption{Sequence similarity between SAbDab train/test set.}
  \label{fig:seq_sim}
  \vspace{-4mm}
\end{figure}

The sequence similarity distribution between the training set and test set is illustrated in Figure \ref{fig:seq_sim}. As we can observe, the average sequence similarity for each CDR region in the training and test set is around 0.3 to 0.5, which shows that there is no potential data leakage issue in this data split strategy.
In addition, we utilize a T-cell receptor dataset released in the structural T-cell receptor database \citep{leem2018stcrdab} to construct a test set with $5,111$ receptors, referred to as STCRDab.
To evaluate the model efficiency, we utilize $5,000$ predicted CDR-H3 loops from the Observed Antibody Space (OAS) \citep{olsen2022oas}, denoted as OAS-H3. Redundant CDR loops are removed from the test set. Statistics of these datasets are listed in Table \ref{tab:datasets}.

{\bf Labels.}
PyIgClassify cluster labels \citep{north2011fccc,adolf2015fccc} are employed as ground-truth labels to assess the retrieval performance of antibody CDR regions. For each PDB structure containing an identified antibody heavy or light chain, PyIgClassify categorizes the conformations of CDRs using a three-tier strategy: chain and position, length, and the similarity of dihedral angles. For instance, the cluster ID L1-10-1 denotes a CDR-L1 with a length of 10 amino acids, where the subcluster 1 is determined based on the similarity of dihedral angles using the affinity propagation clustering method \citep{frey2007clustering}.

\input{section/tab/tab-1-dataset}

\section{Implementation Details}
\label{sec:appendix-param}
In this section, we introduce the implementation details of our {\igseek}.
The MEGNN model introduced in Section \ref{sec:method} consists of three key learnable functions:
\begin{itemize}[topsep=0.5mm, partopsep=0pt, itemsep=0pt, leftmargin=10pt]
    \item The edge module $\phi_e$ (refer to Eq. \ref{eq:megnn:edge}) consists of a two-layer MLP with two Leaky Rectified Linear Unit (LeakyReLU) activation functions \citep{xu2015leakyrelu}. Besides, a dropout function \citep{srivastava2014dropout} with $0.1$ dropout rate is employed on the output of $\phi_e$: 
    $$
    \begin{aligned}
        &\text{CONCAT(Features)} \rightarrow \text{Input} \rightarrow \{ \text{LinearLayer()} \rightarrow \text{LeakyReLU()} \rightarrow \text{LinearLayer()} \\
        &\rightarrow \text{LeakyReLU()} \} \rightarrow 
        \text{Dropout} \rightarrow \text{Output}.
    \end{aligned}
    $$
    \item The coordinate module $\phi_X$ (refer to Eq. \ref{eq:megnn:coordinate}) contains a two-layer MLP that shares weights with the MLP in the edge module $\phi_e$. 
    \item The node module $\phi_h$ (refer to Eq. \ref{eq:megnn:feature}) is a two-layer MLP with one LeakyReLU activation function:
    $$    
        \text{CONCAT(Features)} \rightarrow \text{Input} \rightarrow \{ \text{LinearLayer()} \rightarrow \text{LeakyReLU()} \rightarrow \text{LinearLayer()} \} \rightarrow \text{Output}.
    $$
\end{itemize}

In our experiments, we train the MEGNN model in {\igseek} using PyTorch \citep{paszke2019pytorch} with an Adam optimizer \citep{kingma2015adam} on 4 NVIDIA Tesla A100 GPUs. Table \ref{tab:appendix-param} lists the hyperparameters of {\igseek}.

\section{Baselines}
\label{sec:appendix-baseline}

The first category is structure retrieval model:
\begin{itemize}[topsep=0.5mm, partopsep=0pt, itemsep=0pt, leftmargin=10pt]
    \item {\bf FoldSeek} \citep{van2024foldseek} represents tertiary amino acid interactions using 3D interaction (3Di) structural alphabet, achieving 4 to 5 orders of magnitude speed-up compared to traditional iterative or stochastic structure retrieval methods like CE \citep{shindyalov1998ce}, Dali \citep{holm2020dali}, and TM-align \citep{zhang2005tmalign}. Official code is available at: \href{https://github.com/steineggerlab/foldseek}{https://github.com/steineggerlab/foldseek}.
\end{itemize}

\begin{figure}[t]
  \centering
  \includegraphics[width=0.95\textwidth]{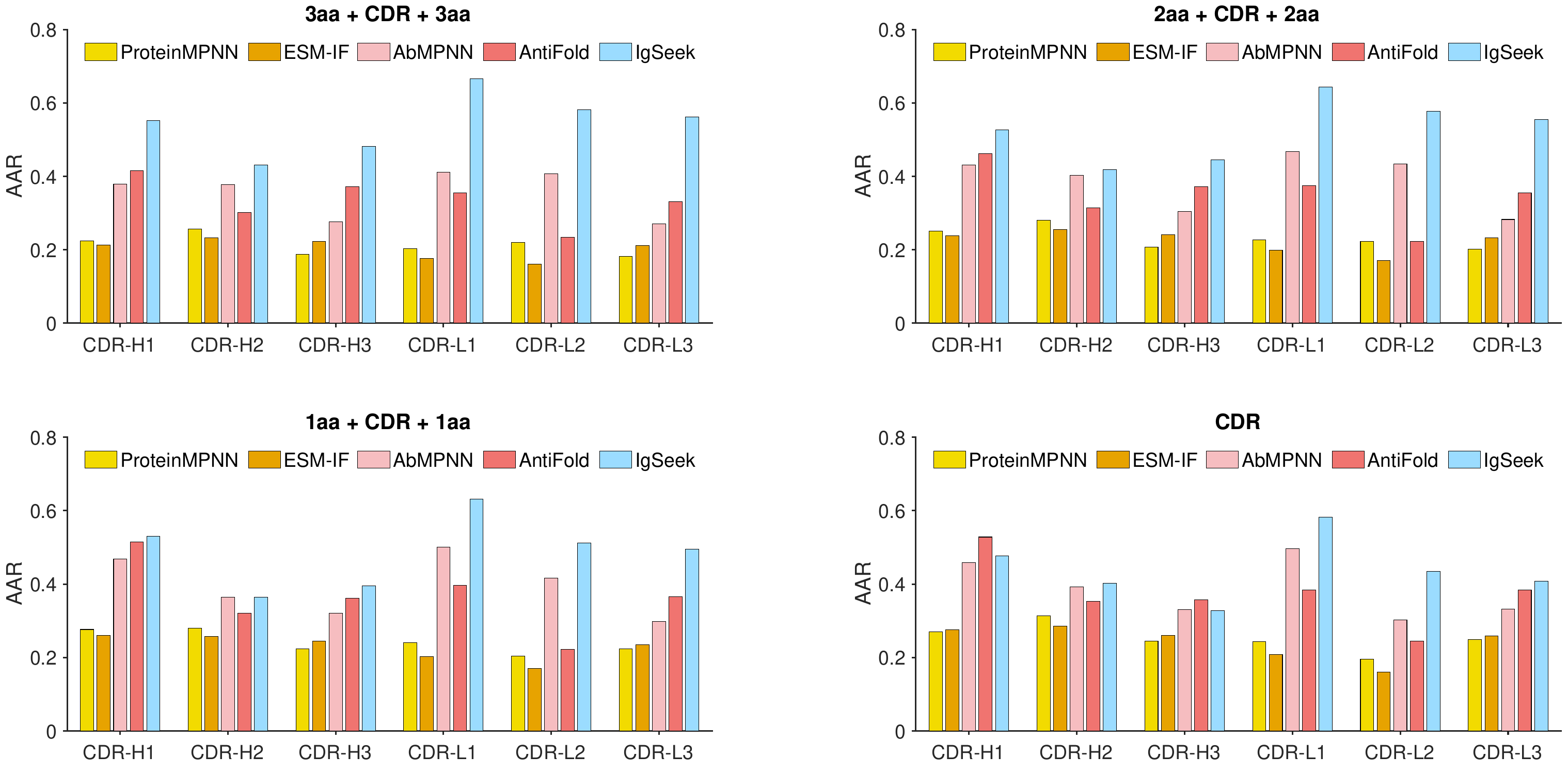}
  \vspace{-2mm}
  \caption{The Comparison of Average AAR on the SAbDab-2024 Dataset using CDRs with extensions of 1 to 3 amino acids on each side in the flanking regions.}
  \label{fig:exp-extension}
\end{figure}

The second category is protein and antibody design models:
\begin{itemize}[topsep=0.5mm, partopsep=0pt, itemsep=0pt, leftmargin=10pt]
    \item {\bf ProteinMPNN}  \citep{dauparas2022mpnn} 
   is a deep learning–based method for protein sequence design that excels in both in silico and experimental evaluations, achieving a sequence recovery of 52.4\% on native protein backbones, compared to 32.9\% for Rosetta \citep{adolf2018rosetta,baek2021rosetta}. By leveraging a message-passing neural network with enhanced input features and edge updates, ProteinMPNN is capable of designing monomers, cyclic oligomers, protein nanoparticles, and protein-protein interfaces, rescuing previously failed designs generated by Rosetta \citep{baek2021rosetta} or AlphaFold \citep{jumper2021alphafold}. 
    Official code is available at: \href{https://github.com/dauparas/ProteinMPNN}{https://github.com/dauparas/ProteinMPNN}.
    \item {\bf ESM-IF1} \citep{hsu2022esmif1} 
    employs a sequence-to-sequence Transformer to predict protein sequences from backbone atom coordinates, which is pre-trained on structures of 12M protein sequences. It achieves 51\% native sequence recovery and 72\% for buried residues.
    Official code is available at: \href{https://github.com/facebookresearch/esm/tree/main/examples/inverse\_folding}{https://github.com/facebookresearch/esm/tree/main/examples/inverse\_folding}.
    \item {\bf AbMPNN} \citep{dreyer2023abmpnn} fine-tunes ProteinMPNN on the SAbDab \citep{dunbar2013sabdab,schneider2021sabdab} dataset for antibody design, outperforming generic protein models in sequence recovery and structure robustness, especially for the hypervariable CDR-H3 loop. The profile of model weights is available at: \href{https://zenodo.org/records/8164693}{https://zenodo.org/records/8164693}.
    \item {\bf AntiFold} \citep{hoie2024antifold} is an antibody-specific inverse folding model fine-tuned from ESM-IF1 \citep{hsu2022esmif1} on solved antibody structures from the SAbDab dataset \citep{dunbar2013sabdab,schneider2021sabdab} and predicted antibody structures from the OAS dataset \citep{kovaltsuk2018oas, olsen2022oas}.
    AntiFold excels in sequence recovery and structural similarity while also demonstrates stronger correlations in predicting antibody-antigen binding affinity in a zero-shot manner.
    Official code is available at: \href{https://github.com/oxpig/AntiFold}{https://github.com/oxpig/AntiFold}.
\end{itemize}

\section{Additional Experiments}
\label{sec:appeidix-exp}
{\bf CDR with extensions.}
In this set of experiments, we compare {\igseek} with protein and antibody design baselines using the SAbDab-2024 dataset. We focus on CDRs with backbone extensions of $n$ amino acids on each side in the flanking regions. Fig. \ref{fig:exp-extension} illustrates the results for varying values of $n = 0, 1, 2, 3$. As we can observe, the performance of {\igseek} improves with the inclusion of additional amino acids in the given structure, , which aligns with the fact that more input structural information can be encoded into the CDR representation. In contrast, other baseline models are adversely affected by hallucinations stemming from conserved backbone structures. Notably, when $n=3$, {\igseek} consistently outperforms its competitors by at least $5\%$ and $18\%$ for heavy chain and light chain CDR loops, respectively. This further demonstrates that the retrieval-based strategy employed by {\igseek} effectively mitigates hallucinations during CDR sequence generation.

\input{section/tab/tab-4-K}

{\bf Influence of value $K$.}
In this set of experiments, we conduct experiments on the SAbDab-2024 dataset to evaluate the impact of varying parameter $K$ in {\igseek}. Table \ref{tab:param-k} reports the average AAR of {\igseek} across different values of $K$ on the SAbDab-2024 dataset. As we can observe, the performance of IgSeek exhibits a decline as $K$ increases. In our implementation, we set $K=10$ rather than $5$ as {\igseek} achieves comparable results while preserving enhanced sequence diversity.

%% file: section/2-related-work.tex
\section{Related Work}
\label{sec:related-work}

{\bf Protein Structure Retrieval.} With the growth of the volume of protein structures, structure retrieval has become a critical task in protein data management. 
AlphaFind \citep{prochazka2024alphafind} is a web tool designed to identify structurally similar proteins in AlphaFold Database \citep{varadi2022alphafolddb} by compressing data from $\sim$23 TB to $\sim$20 GB using vector embeddings, narrowing down candidates with a neural network. The similarity of the search result is evaluated by US-align~\citep{zhang2022usalign}. Another state-of-the-art method, FoldSeek \citep{van2024foldseek}, accelerates protein structure searches by representing tertiary amino acid interactions as sequences over a 3D interaction structural alphabet, which derives from vector quantization by VQ-VAE \citep{oord2017vqvae}. 
However, the representation only models the structure of two contiguous residues in a chain.  

\input{section/tab/tab-3-settings}

{\bf Protein Inverse Folding.} Protein inverse folding aims to predict diverse sequences that can fold into a given protein structure.
ProteinMPNN \citep{dauparas2022mpnn} is a deep learning–based method for protein sequence design that excels in both in silico and experimental evaluations. By leveraging a message-passing neural network with enhanced input features and edge updates, ProteinMPNN is capable of designing monomers, cyclic oligomers, protein nanoparticles, and protein-protein interfaces, rescuing previously failed designs generated by Rosetta \citep{adolf2018rosetta,baek2021rosetta} or AlphaFold \citep{jumper2021alphafold}.
ESM-IF1 \citep{hsu2022esmif1} employs a sequence-to-sequence Transformer to predict protein sequences from backbone atom coordinates. 

{\bf Antibody Inverse Folding.} AbMPNN \citep{dreyer2023abmpnn} inherits the model architecture of ProteinMPNN, and trains an antibody-specific variant for antibody design. It outperforms generic protein design models in sequence recovery and structure robustness, especially for hyper-variable CDR-H3 loops. AntiFold \citep{hoie2024antifold} is an antibody-specific inverse folding model, which is fine-tuned on ESM-IF1, with both solved and predicted antibody structures. However, it should be emphasized that Antifold infers CDR sequences based on the structure of the variable domain and the sequence of the framework regions. Consequently, the accuracy of CDR sequence inference is influenced not only by the structure of the CDRs but also by the sequence and structural information of the framework regions. Previous studies utilizing antibody sequence language models without structural information have demonstrated that the sequence of the framework regions can partially predict the CDR sequences, particularly for relatively conserved residues. As a result, the requirement for the framework sequence as input complicates the inference of CDR sequences that can bind to different antigens while maintaining an identical framework.

{\bf Antibody Co-Design.} In recent years, deep learning models have emerged as powerful data-driven approaches for antibody design. RefineGNN \citep{jin2022refinegnn} is the first structure sequence co-design method that alternatively predicts the atom coordinates and residue types in CDRs by auto-regression. DiffAb \citep{luo2022diffab} and IgGM \citep{wang2024iggm} utilize diffusion models to generate the structure and sequence of CDRs based on the framework regions and the target antigen, with DiffAb oriented for specific antigens. MEAN \citep{kong2023mean} and dyMEAN \citep{kong2023dymean} employ graph neural networks to predict the structure and sequence of CDRs. Table \ref{tab:settings} presents a comparative analysis of various antibody design task configurations.

%% file: section/tab/tab-3-settings.tex
\begin{table*}[t]
\small
\caption{Settings of different antibody (Ab) design tasks.}
\vspace{-3mm}
\centering
\label{tab:settings}
\begin{tabular}{cccc|cc}
\toprule
\multirow{2}{*}{Category} & \multicolumn{3}{c|}{Input} &  \multicolumn{2}{c}{Output} \\
\cmidrule{2-6}
& Ab Framework & Ab CDR & Antigen & CDR Structure & CDR Sequence\\
\midrule
Antibody Inverse Folding & \Checkmark & \Checkmark & \XSolidBrush & \XSolidBrush & \Checkmark \\
\midrule
Antibody Co-design & \Checkmark & \Checkmark & \Checkmark & \Checkmark & \Checkmark \\
\midrule
Sequence Design (ours) & \XSolidBrush & \Checkmark & \XSolidBrush & \XSolidBrush & \Checkmark \\
\bottomrule
\end{tabular}
\vspace{-4mm}
\end{table*}

%% file: section/3-preliminary.tex
\section{Preliminaries and Problem Formulation}
\label{sec:preliminary}

\begin{figure}[h]
  \centering
  \includegraphics[width=0.85\textwidth]{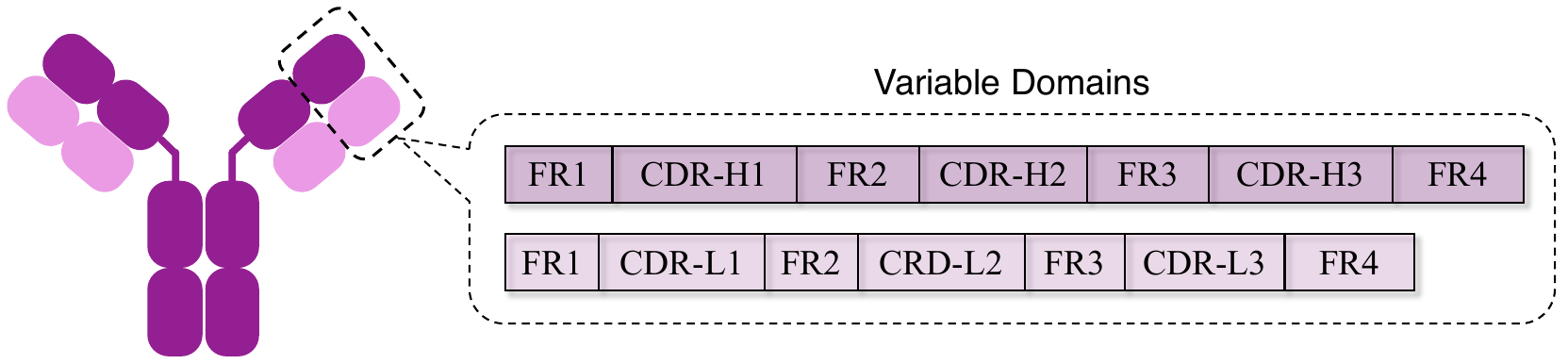}
  \caption{Antibody Structure}
  \label{fig:antibody}
\end{figure}

Antibodies are sophisticated Y-shaped protein characterized by their distinctive structural architecture, comprising two identical sets of polypeptide chains. Each set consists of a heavy chain and a light chain, with both chains exhibiting a modular organization of constant and variable regions. While the constant regions maintain high sequence conservation across different antibody molecules, the variable regions display significant diversity, enabling specific antigen recognition and binding capabilities. Within the variable domains, the structural organization follows a precise pattern defined by the IMGT numbering scheme, alternating between four framework regions (FRs) and three complementarity determining regions (CDRs). The FRs provide the structural scaffold, while the CDRs form the antigen-binding interface and are primarily responsible for the specificity and affinity of antigen recognition. These hypervariable CDR loops represent the most critical determinants of antibody-antigen interactions and are consequently the primary focus of rational antibody engineering and design efforts.

We represent the 3D structure of a CDR as a geometric graph $G=(V,E)$ with node set $V$ and edge set $E$ \citep{jing2020gvp,jin2022refinegnn,zhang2023gearnet}. Each node $v_i \in V$ denotes an amino acid residue, associated with a multi-channel 3D coordinate matrix $\vect{X}_i \in \mathbb{R}^{c \times 3}$, where $c$ is the channel size, i.e., the number of atoms in the residue $v_i$. In this paper, we consider the four backbone atoms \{N, C$_\alpha$, C, O\} that are independent to residue type, i.e., $c = 4$. Each edge $e_{ij} \in E$ denotes an interaction between $v_i$ and $v_j$, if the Euclidean distance between their C$_\alpha$ atoms is within a threshold $\theta$. The neighborhood of a node $v_i$, denoted as $\mathcal{N}_{i}$, consists of the adjacency nodes of $v_i$, that is, $\{v_j | (v_i, v_j)\in E\}$.

{\bf CDR Sequence Design.} Given the structure $G=(E, V)$ of a CDR and the multi-channel 3D coordinate of each residue, in this paper, we aim to reconstruct the corresponding sequence of the CDR, denoted as $\vect{s} = \{s(i) | i \in [1, \cdots, |V| ]\}$, where $s(i)$ is the amino acid type of residue $v_i$.

E(3) Equivalence is an important property in modeling the 3D structures \citep{fuchs2020se3transformer,batzner2022nequip,liao2023equiformer}. Formally, let $\mathcal{X}$ and $\mathcal{Y}$ be two vector spaces, with $T_{\mathcal{X}}(g): \mathcal{X} \rightarrow \mathcal{X}$ and $T_{\mathcal{Y}}(g): \mathcal{Y} \rightarrow \mathcal{Y}$ representing two sets of transformations for the abstract group $g \in E(3)$. A function $\phi: \mathcal{X} \rightarrow \mathcal{Y}$ is E(3) Equivariant to $g$ if it satisfies the following condition: 
\begin{align}
    \phi(\{T_{\mathcal{X}}(g)\vect{x}_i, \vect{h}_i \}_{i=1}^n) = T_{\mathcal{Y}}(g) \phi(\{ \vect{x}_i, \vect{h}_i \}_{i=1}^n),
\end{align}
where $\vect{x}_{i} \in \mathbb{R}^3$ denotes the input 3D coordinates and $\vect{h}_i \in \mathbb{R}^{d}$ is the $d$-dimensional features of a node, respectively. This inductive bias guarantees that $\phi$ preserves equivariant transformation regarding transformation of the coordinate system in E(3) group \citep{satorras2021egnn,huang2022gmn,liao2023equiformer}. A typical example for this transformation operation in the space $\mathcal{X}$ is given by $T_{\mathcal{X}}(g)\vect{x}_i^{(0)} = \vect{R}\vect{x}_i^{(0)} + \vect{b}$, where $\vect{R}\in \mathbb{R}^{3 \times 3}$ is an orthogonal matrix and $\vect{b}$ is the bias term.

To achieve equivalence, equivariant graph neural networks are proposed~\citep{satorras2021egnn, huang2022gmn, kong2023dymean, kong2023mean}, which follows a general message-passing framework as shown in Eq.~\ref{eq:egnn:message}-\ref{eq:egnn:coordinate}.
Here, $\vect{m}_{j \rightarrow i}^{(l)}$ denotes the messages propagated from node $v_j$ to $v_i$, and $d_{ij}^{(l-1)} = dist(v_i, v_j)$ denotes the Euclidean distance between $v_i$ and $v_j$, and $\vect{x}_{ij}^{(l-1)}$ denotes coordinate differences between $v_i$ and $v_j$ at the $(l-1)$-th layer.
\begin{align}
    \vect{m}_{j \rightarrow i}^{(l)} &= \psi_1 \left( \vect{h}_i^{(l-1)}, \vect{h}_j^{(l-1)}, \vect{x}_{ij}^{(l-1)}, d_{ij}^{(l-1)} \right), \label{eq:egnn:message} \\
    \vect{h}_i^{(l)} &= \psi_2 \left(\vect{h}_i^{(l-1)}, \sum_{v_j \in \mathcal{N}_i} \vect{m}_{j \rightarrow i}^{(l)} \right), \label{eq:egnn:feature} \\
    \vect{x}_i^{(l)} &= \psi_3 \left( \vect{x}_i^{(l-1)}, \vect{x}_{ij}^{(l-1)} \sum_{j} \psi_4 \left( \sum_{v_j \in \mathcal{N}_i} \vect{m}_{j \rightarrow i}^{(l)} \right) \right). \label{eq:egnn:coordinate}
\end{align}
The functions $\{\psi_1, \psi_2,\psi_3, \psi_4\}$ are equivariant transformations, typically implemented as Multi-Layer Perceptrons (MLPs) to leverage the universal approximation \citep{funahashi1989mlp,cybenko1989mlp,hornik1991mlp}. In this process, the feature $\vect{h}_i^{(l)}$ remains E(3) invariant, while the coordinate $\vect{x}_i^{(l)}$ is E(3) equivariant.

%% file: section/4-method.tex
\section{Methodology: IgSeek}
\label{sec:method}

In this section, We present IgSeek, a retrieval-based framework for CDR sequence design. The core of IgSeek is isomorphic structure retrieval from a comprehensive antibody CDR database. As shown in Fig.~\ref{fig:framework}, IgSeek first builds a vector database using a pre-trained Multi-channel Equivariant Graph Neural Network (MEGNN) to encode CDR structural information. For a target CDR structure G, IgSeek generates its embedding using MEGNN, retrieves K-nearest structurally similar CDR loops from the database, and predicts G's sequence through ensemble and Bernoulli sampling of the retrieved sequences. In the following, we will present the model design of the MEGNN encoder in Section~\ref{sec:subsec-encoder}, discuss the learning objective and the sequence prediction in Section~\ref{sec:subsec-decoder}, followed by model analysis in Section~\ref{sec:subsec-thm}.

\subsection{Multi Channel Equivariant Encoder}
\label{sec:subsec-encoder}

Recall that each amino acid residue $v_i$ is represented by its four backbone atoms, thereby we extend the general single-channel EGNN layer~\citep{satorras2021egnn,huang2022gmn} to a multi-channel layer, with each channel corresponding to a specific atom. Unlike existing approaches~\citep{kong2023dymean, hoie2024antifold} that leverage domain knowledge of the well-conserved antibody backbone structure, our MEGNN encoder generates CDR embeddings exclusively based on the antibody CDR structure, without relying on any prior backbone knowledge.

For a 3D CDR structure $G$, the MEGNN encoder takes the initial features of each residue $v_i$, denoted as $\vect{h}_i^{(0)} \in \mathbb{R}^{ d}$, along with the perturbed coordinates $\vect{\hat{X}}_i \in \mathbb{R}^{c \times 3}$ as input. Here, $c$ denotes the number of atoms, which is set to $c=4$, and $\vect{h}_i^{(0)}$ is initialized by a uniform distribution. 
$\vect{\hat{X}}_i = \vect{X}_i + \mathcal{N}(0, \sigma)$, where $\mathcal{N}(0, \sigma)$ denotes a small Gaussian noise. This perturbation introduces variability that enhances the robustness of the model.  

{\bf Multi-channel Equivariant Message Passing.}
The $l$-th layer of MEGNN updates both the node features $\vect{h}_i^{(l)}$ and coordinates $\vect{X}_i^{(l)}$ by Eq. \ref{eq:megnn:distance}-\ref{eq:megnn:feature}, where $\rho$ is a distance computation function, $\phi_e$, $\phi_X$ and $\phi_h$ are neural network transformations. The update process is defined as follows:
\begin{align}
    \vect{X}_{ij}^{(l-1)}, \vect{z}_{ij}^{(l-1)} &= \rho \left( \vect{X}_i^{(l-1)}, \vect{X}_j^{(l-1)}, e_{ij} \right), \label{eq:megnn:distance} \\
    \vect{h}_{e_{ij}}^{(l)} &= \phi_e \left( \text{CONCAT}\left(\vect{h}_i^{(l-1)}, \vect{h}_j^{(l-1)}, \vect{z}_{ij}^{(l-1)} \right) \right), \label{eq:megnn:edge} \\
    \vect{X}_i^{(l)} &= \phi_X \left( \vect{X}_i^{(l-1)}, \{ \vect{h}_{e_{ij}}^{(l)}, \vect{X}_{ij}^{(l-1)} | v_j \in \mathcal{N}_{i} \} \right), \label{eq:megnn:coordinate}\\
    \vect{h}_i^{(l)} &= \phi_h \left( \vect{h}_i^{(l-1)}, \{ \vect{h}_{e_{ij}}^{(l)} | v_j \in \mathcal{N}_i  \} \right). \label{eq:megnn:feature}
\end{align}

Specifically, MEGNN first computes the coordinate differences $\vect{X}_{ij}^{(l-1)}$ and the square distance $\vect{z}_{ij}^{(l-1)}$ between each pair of backbone atoms among different residues in $\rho$ (Eq.~\ref{eq:megnn:distance}) as below: 
\begin{align}
    \vect{X}_{ij}^{(l-1)} = \vect{X}_i^{(l-1)} - \vect{X}_j^{(l-1)}, \quad \vect{z}_{ij}^{(l-1)} = (\vect{X}_{ij}^{(l-1)})^{\top} \vect{X}_{ij}^{(l-1)}. \nonumber
\end{align}
Subsequently, an edge module $\phi_e$ generates the edge feature $\vect{h}_{e_{ij}}^{(l)}$ for each edge $e_{ij} = (v_i, v_j) \in E$. In Eq.~\ref{eq:megnn:edge}, the node features of $v_i$ and $v_j$, i.e., $\vect{h}_i^{(l-1)}$, $\vect{h}_j^{(l-1)}$, along with the fattened coordinate difference $(\vect{z}_{ij}^{(l-1)})$, are concatenated and transformed by an MLP, generating the output edge feature for the $l$-th layer. Next, the coordinate module $\phi_X$ updates the node coordinates $\vect{X}_i^{(l)}$ using the updated edge feature $\vect{h}_{e_{ij}}^{(l)}$ and the coordinate differences $\vect{X}_{ij}^{(l-1)}$ in Eq.~\ref{eq:megnn:coordinate}. Specifically, for each node $v_i$, $\phi_X$ first computes the message $\vect{m}_{j \rightarrow i}$ propagated from its neighbor $v_j$, and then updates the coordinates $\vect{X}_i^{(l)}$ of $v_i$ by aggregating the messages from its neighborhood:
\begin{align}
    \vect{m}_{j \rightarrow i} &= \text{MLP}\left( \vect{h}_{e_{ij}}^{(l)} \right) \cdot \vect{X}_{ij}^{(l-1)}, \\ \vect{X}_i^{(l)} &= \vect{X}_i^{(l-1)} + \frac{1}{|\mathcal{N}_i|} \sum_{v_j \in \mathcal{N}_i} \vect{m}_{j \rightarrow i}.
\end{align}
Finally, the node module $\phi_h$ updates the node representation $\vect{h}_i^{(l)}$ by Eq.~\ref{eq:megnn:feature}. For each node $v_i$, $\phi_h$ aggregates the features of the adjacent edges into $\vect{h}_{agg_i}^{(l)}$ and combines the node representation $\vect{h}_i^{(l-1)}$ from the $(l - 1)$-th layer with the aggregated feature using a residual connection \citep{he2016resnet}:
\begin{align}
    \vect{h}_{agg_i}^{(l)} &= \sum_{j \in \mathcal{N}_i} { \vect{h}^{(l)}_{e_{ij}}}, \\ \quad \vect{h}_{i}^{(l)} &= \vect{h}_i^{(l-1)} + \text{MLP}\left( \text{CONCAT}(\vect{h}_i^{(l-1)}, \vect{h}_{agg_i}^{(l)}) \right). \nonumber
\end{align}

{\bf CDR Embedding Generation.} After the equivariant message passing through an $L$-layer MEGNN, we employ a $\text{READOUT}$ function to aggregate the final node features to generate the representation of CDR $G$ that consists of $n$ nodes (amino acids) as Eq.~\ref{eq:megnn:readout},
\begin{align}
\label{eq:megnn:readout}
    \vect{h}_G = \text{READOUT}(\{\vect{h}_i^{(L)}\}_{i=1}^{n}).
\end{align}
The READOUT function can be a permutation invariant function, e.g. summation and element-wise mean pooling functions. In our implementation, we set the READOUT function as element-wise mean pooling by default. Algorithm \ref{alg:megnn} summarizes the forward pass of MEGNN.

\input{section/alg/alg-1-megnn}

\subsection{Learning Objective and Sequence Generation}
\label{sec:subsec-decoder}
We train the MEGNN encoder by a self-supervised distance prediction task that explicitly aligns pairs of similar CDR in a given database. The goal is to align the structural representation of similar CDR pairs. 
For a CDR database $\mathcal{B} = \{G_1, G_2, \cdots, G_n\}$, we construct a training dataset $\mathcal{T} = \{(G_i, G_j), \cdots\}$ containing pairs of fixed-length CDRs whose TM-Score, calculated by TM-align \citep{zhang2005tmalign}, exceeds a specified threshold. Given a pair of CDRs $(G_i, G_j)$, we first generate their representations using MEGNN, denoted as $\vect{h}_{G_i}$ and $\vect{h}_{G_j}$, respectively.
Next, we predict the {\em Root Mean Square Deviation (RMSD)} of the two CDR structures by feeding the concatenation of $\vect{h}_{G_i}$ and $\vect{h}_{G_j}$ into an MLP decoder as Eq.~\ref{eq:megnn:distance_pred}:
\begin{align}
\label{eq:megnn:distance_pred}
    \widehat{d}(G_i, G_j) = \text{MLP} \left(\text{CONCAT} \left(\vect{h}_{G_i}, \vect{h}_{G_j} \right) \right). 
\end{align}

{\bf Loss Function.} 
The learning objective is to minimize the Mean Square Error between the predicted distance $\widehat{d}(G_i, G_j)$ and the actual distance $d(G_i, G_j)$ in the training dataset $\mathcal{T}$: 
\begin{align}
\label{eq:megnn:loss}
    \mathcal{L} = \frac{1}{|\mathcal{T}|}\sum_{(G_i, G_j) \in \mathcal{T} } \|\widehat{d}(G_i, G_j) - d(G_i, G_j) \|^2. 
\end{align}
Here, the actual distance $d(G_i, G_j)$ is computed as the RMSD of the two CDRs for their backbone atoms.
Since we do not have prior knowledge of the CDR cluster labels, our approach can be interpreted as an unsupervised geometric learning model. By minimizing the loss function defined in Eq. \ref{eq:megnn:loss}, the model effectively generates CDR embeddings that reflect the structural relationships among the CDRs in the dataset.

{\bf CDR Sequence Generation.}
Once the model training is complete, we establish a CDR vector database $\mathcal{Z}$, where each CDR$_i$ is represented by a triplet $(\vect{s}_i, G_i, \vect{h}_{G_i})$ consisting of amino acid sequence $\vect{s}_i$, its backbone structure graph $G_i$ and its embedding $\vect{h}_{G_i}$ generated by the MEGNN encoder via Eq.~\ref{eq:megnn:readout}. IgSeek is then able to infer the amino acid sequence of a CDR by querying its backbone structure in the database $\mathcal{Z}$. Let $\vect{s}_q$ denote the query CDR sequence with a length of $L$. At each position $l \in \{1,  \cdots, L\}$, the residue $\vect{s}_q(l)$ is selected from one of the $20$ amino acids, denoted as $a_i$ for  $ i \in \{ 1, \cdots, 20\}$. Then, the inference of the CDR sequence $\vect{s}_q$ given its backbone structure $G_q$ follows four steps: 
{\em (i)} first, the MEGNN encoder generates the embedding of $G_q$, denoted as $\vect{h}_{G_q}$. 
{\em (ii)} Second, the embedding $\vect{h}_{G_q}$ is used as the search key to perform a $K$-NN search in the database $\mathcal{Z}$, obtaining a set of $K$ CDRs of equal length $L$, denoted as $\mathcal{Z}_q = \{(\vect{s}_1, G_1, \vect{h}_{G_1}), (\vect{s}_2, G_2, \vect{h}_{G_2}), \cdots, (\vect{s}_K, G_K, \vect{h}_{G_K})\}$. 
{\em (iii)} Given the $K$ sequences $\mathcal{S}_q = \{\vect{s}_1, \cdots, \vect{s}_K\}$, we derive the probability of amino acid $a_i$ occurring at position $l$ of the predicted sequence $\hat{\vect{s}}_q$ as follows:
\begin{align}
    p\left(\hat{\vect{s}}_q(l) = a_i| S_q\right) = \frac{1}{K} \sum_{s_k \in \mathcal{S}_q } \mathbb{I}(\vect{s}_k(l), a_i), \nonumber
\end{align}
where $\mathbb{I}(\vect{s}_k(l), a_i) \in \{ 0, 1\}$ is a binary indicator that equals $1$ if the amino acid $a_i$ occurs at the position $l$ of sequence $\vect{s}_k$, and $0$ otherwise. {\em (iv)} To derive the final inferred sequence $\hat{s}_q$, we sample the amino acid at each position $l$ according to the generated probability distribution:
\begin{align}
    \hat{\vect{s}}_q(l) \sim p\left(\hat{\vect{s}}_q(l) | \mathcal{S}_q\right). \nonumber
\end{align}
Algorithm \ref{alg:train} outlines the training process, and Algorithm \ref{alg:design} presents the antibody CDR sequence design process, respectively.

\input{section/alg/alg-2-train}
\input{section/alg/alg-3-igseek}

\subsection{Analysis}
\label{sec:subsec-thm}
{\bf Model Complexity.} Given a 3D CDR structure represented by $G = (V,E)$, the initialized coordinates, node features, and the graph structure contribute a space complexity of $O(|V| \cdot d + |V| \cdot c + |E|) = O(|V|\cdot d + |E|)$, where $d$ denotes the hidden dimension of features and $c$ denotes the channel size. In MEGNN, the space complexity is dominated by the edge features, which have a complexity of $O(|E| \cdot d)$, and square distance $\vect{z}$ with a complexity of $O(|E| \cdot c^2)$. Consequently, the overall space complexity is $O(|E|\cdot d)$, which is linear to the input graph size.
Regarding the computational complexity of MEGNN, the dominant component is the edge module $\phi_e$ introduced in Eq.~\ref{eq:megnn:edge}, which has a time complexity of $O(|E| \cdot (2d+3c)^2 + |E| \cdot d^2 + 3c) = O(|E| \cdot d^2)$.

{\bf Coordinate Equivariance and Representation Invariance.} The following theorem shows that MEGNN is E(3) equivariant with respect to the initial coordinate $\vect{X}_i^{(0)}$ and E(3) invariant with respect to the representations $\vect{h}$ of the input CDR, respectively.
\begin{theorem}
\label{thm:thm1-equivariance}
    For any transformation $g \in E(3)$, we have $\vect{h}_i, T_{\mathcal{Y}}(g) \vect{X}_i^{(L)} = \text{MEGNN} \left( \vect{h}_i^{(0)}, T_{\mathcal{X}}(g) \vect{X}_i^{(0)}, G \right)$, where $T_{\mathcal{X}}$ and $T_{\mathcal{Y}}$ 
    $:= \vect{R} \vect{X} + \vect{b}$ denotes the transformation of $\vect{X}$ in the input space $\mathcal{X}$ (resp. output space $\mathcal{Y}$), $\vect{R}$ is an orthogonal matrix, and $\vect{b}$ is the bias.
\end{theorem}
The theorem indicates that MEGNN can be generalized to arbitrary E(3) group operations (refer to Section~\ref{sec:preliminary}), which showcases the data efficiency of MEGNN.

\begin{proof}
    We assume that $\vect{h}_i^{(0)}$ is invariant to E(3) transformation operations on the coordinate $\vect{X}_i^{(0)}$, since $\vect{h}_i^{(0)}$ is generated from uniform distribution and no absolute information of $\vect{X}_i^{(0)}$ is encoded into $\vect{h}_i^{(0)}$. 
    Then, for the E(3) transformation $g:= \vect{R} \vect{X} + \vect{b}$, where orthogonal matrix $\vect{R}\in O(3)$ and bias $\vect{b} \in \mathbb{R}^3$, we have:
    $$
    \begin{aligned}
        \vect{R} \vect{X}_{i}^{(l-1)} + \vect{b} - (\vect{R} \vect{X}_{j}^{(l-1)} + \vect{b}) &= \vect{R} \vect{X}_{ij}^{(l-1)}, \\
        (\vect{R} \vect{X}_{ij}^{(l-1)} )^{\top} \vect{R} \vect{X}_{ij}^{(l-1)} &= z_{ij}^{(l-1)}.
    \end{aligned}
    $$
    Therefore, the output $z_{ij}^{(l-1)}$ of Eq. \ref{eq:megnn:distance} is E(3) invariant to transformation $g$.
    
    As for Eq. \ref{eq:megnn:edge}, since $\vect{h}_i$, $\vect{h}_j$, and $z_{ij}^{(l-1)}$ are invariant to E(3) transformation operations, we can derive that $\vect{h}_{e_{ij}}^{(l)}$ is E(3) invariant.
    
    Next, we will prove Eq. \ref{eq:megnn:coordinate} is E(3) equivariant. 
    $$
    \begin{aligned}
        \vect{R} \vect{X}_i^{(l-1)} + \vect{b} + \frac{1}{|\mathcal{N}_i|} \sum_{v_j \in \mathcal{N}_i} \text{MLP} \left( \vect{h}_{e_{ij}}^{(l)} \right) \cdot \vect{R} \vect{X}_{ij}^{(l-1)}  &= \vect{R} \left(\vect{X}_i^{(l-1)} + \frac{1}{|\mathcal{N}_i|} \sum_{v_j \in \mathcal{N}_i} \vect{m}_{j \rightarrow i} \right) + \vect{b} \\
        &= \vect{R} \vect{X}_i^{(l)} + \vect{b}.
    \end{aligned}
    $$
    Therefore, we have proven that any E(3) transformation operations on $\vect{X}_i^{(l-1)}$ leads to the same E(3) transformation operations on $\vect{X}_i^{(l)}$ using Eq. \ref{eq:megnn:coordinate}.
    
    Finally, it is easy to verify that Eq. \ref{eq:megnn:feature} is E(3) invariant as $\vect{h}_i^{(l-1)}$ and $\vect{h}_{e_{ij}}^{(l)}$ are E(3) invariant. 
    
    In conclusion, for an $L$-layer MEGNN model, any transformation $g \in E(3)$ on the input coordinate $\vect{X}^{(0)}$ will lead to the same E(3) transformation operations on the output coordinate $\vect{X}^{(L)}$ while the representations $\vect{h}^{(L)}$ still remain E(3) invariant:
    $$
        \vect{h}_i, T_{\mathcal{Y}}(g) \vect{X}_i^{(L)} = \text{MEGNN} \left( \vect{h}_i^{(0)}, T_{\mathcal{X}}(g) \vect{X}_i^{(0)}, G \right).
    $$
    This finishes the proof.
\end{proof}

%% file: section/alg/alg-1-megnn.tex
\begin{algorithm}[t]
\caption{Multi-channel Equivariant Graph Neural Network (MEGNN)}\label{alg:megnn}
\begin{algorithmic}
    \STATE {\bfseries Input:} Antibody CDR Structure $G=(V,E)$, initial features $\vect{h}_i^{(0)}$ and coordinates $\vect{X}_i^{(0)}$ for each node $v_i \in V$
    \STATE {\bfseries Output:} Antibody CDR representation $\vect{h}_{G}$
    \STATE Initialize coordinates $\hat{\vect{X}}_i \leftarrow \vect{X}_i + \mathcal{N}(0, \sigma)$
    \FOR{layer $l=1$ {\bfseries to} $L$}
        \FOR{$v_i \in V$}
            \FOR{$v_j \in \mathcal{N}_i$}
                \STATE Calculate the coordinate differences: $\vect{X}_{ij}^{(l-1)} \leftarrow \vect{X}_i^{(l-1)} - \vect{X}_j^{(l-1)}$
                \STATE Calculate the square distance: $\vect{z}_{ij}^{(l-1)} \leftarrow (\vect{X}_{ij}^{(l-1)})^{\top} \vect{X}_{ij}^{(l-1)}$        
                \STATE Update the edge feature: $\vect{h}_{e_{ij}}^{(l)} \leftarrow \phi_e \left( \text{CONCAT}\left(\vect{h}_i^{(l-1)}, \vect{h}_j^{(l-1)}, \vect{z}_{ij}^{(l-1)} \right) \right)$
                \STATE Derive the propagated information: $\vect{m}_{j \leftarrow i} \leftarrow \text{MLP}\left( \vect{h}_{e_{ij}}^{(l)} \right) \cdot \vect{X}_{ij}^{(l-1)}$ 
            \ENDFOR
            \STATE Update the coordinate: $\vect{X}_i^{(l)} \leftarrow \vect{X}_i^{(l-1)} + \frac{1}{|\mathcal{N}_i|} \sum_{v_j \in \mathcal{N}_i} \vect{m}_{j \rightarrow i}$
            \STATE Derive the aggregated edge feature: $\vect{h}_{agg_i}^{(l)} \leftarrow \sum_{j \in \mathcal{N}_i} { \vect{h}^{(l)}_{e_{ij}}}$ 
            \STATE Update the node representation: $\vect{h}_{i}^{(l)} \leftarrow  \vect{h}_i^{(l-1)} + \text{MLP}\left( \text{CONCAT}(\vect{h}_i^{(l-1)}, \vect{h}_{agg_i}^{(l)}) \right)$
        \ENDFOR
    \ENDFOR
    \STATE Generate the representation of the input CDR structure: $\vect{h}_G \leftarrow \text{READOUT}(\{\vect{h}_i^{(L)}\}_{i=1}^{n})$ 
    \STATE {\bfseries Return:} CDR representation $\vect{h}_{G}$
\end{algorithmic}
\end{algorithm}

%% file: section/alg/alg-2-train.tex
\begin{algorithm}[t]
\caption{CDR Vector Database Construction}\label{alg:train}
\begin{algorithmic}
    \STATE {\bfseries Input:} Training set $\mathcal{T} = \{ (G_{i1}, G_{i2}) \}_{i=1}^{|\mathcal{T}|}$, training epoch $T$, CDR database $\mathcal{B}= \{ (\vect{s}_j, G_j) \}_{j=1}^{|\mathcal{B}|}$
    \STATE {\bfseries Output:} CDR vector database $\mathcal{Z}$
    \FOR{$t = 1$ {\bfseries to} $T$}
        \FOR{$i = 1$ {\bfseries to} $|\mathcal{T}|$}
            \STATE Initialize feature matrices $\vect{H}_{i1}$ of $G_{i1}$ and feature matrices $\vect{H}_{i2}$ of $G_{i2}$, respectively
            \STATE Generate graph representation for the $i$-th training CDR pair:
            $\vect{h}_{G_{i1}} \leftarrow \text{MEGNN} (G_{i1}, \vect{H}_{i1}, \vect{X}_{i1}), \quad \vect{h}_{G_{i2}} \leftarrow \text{MEGNN} (G_{i2}, \vect{H}_{i2}, \vect{X}_{i2})$
            \STATE Predict the RMSD between $h_{G_{i1}}$ and $h_{G_{i2}}$: $\widehat{d}(G_{i1}, G_{i2}) \leftarrow \text{MLP} \left(\text{CONCAT} \left(\vect{h}_{G_{i1}}, \vect{h}_{G_{i2}} \right) \right)$
            \STATE Compute the loss function: $\mathcal{L} \leftarrow \frac{1}{|\mathcal{T}|}\sum_{(G_{i1}, G_{i2}) \in \mathcal{T} } \|\widehat{d}(G_{i1}, G_{i2}) - d(G_{i1}, G_{i2}) \|^2$
        \ENDFOR
        \STATE Update the model weights $\vect{W}$ to minimize $\mathcal{L}$ using $\frac{\partial \mathcal{L}}{\partial \vect{W}}$
    \ENDFOR
    \FOR{$j=1$ {\bfseries to} $|\mathcal{B}|$}
        \STATE Generate graph representation $G_j \leftarrow \text{MEGNN} (G_{j}, \vect{H}_{j}, \vect{X}_{j}) $\;
        Add the triplet $(\vect{s}_j, G_j, \vect{h}_{G_j})$ into $\mathcal{Z}$
    \ENDFOR
    \STATE {\bfseries Return:} Vector database $\mathcal{Z}$
\end{algorithmic}
\end{algorithm}

%% file: section/alg/alg-3-igseek.tex
\begin{algorithm}[t]
\caption{Sequence Generation}\label{alg:design}
\begin{algorithmic}
    \STATE {\bfseries Input:} Query structure $G_q$, MEGNN $\phi$, CDR vector database $\mathcal{Z}$
    \STATE {\bfseries Output:} Predicted sequence $\hat{\vect{s}}_q$
    \STATE Initialize feature matrix $\vect{H}_q$ and coordinates $\vect{X}_q$
    \STATE Generate graph representation $G_q \leftarrow \text{MEGNN} (G_{q}, \vect{H}_{q}, \vect{X}_{q})$
    \STATE Retrieve the $K$-nearest neighbors of $\vect{h}_{G}$ in the database $\mathcal{Z}$ as $\mathcal{Z}_q$
    \STATE Derive the probability of amino acid $a$ at the $l$-th position: $p\left(\hat{\vect{s}}_q(l) = a_i| S_q\right) = \frac{1}{K} \sum_{s_k \in \mathcal{S}_q } \mathbb{I}(\vect{s}_k(l), a_i)$
    \STATE Sample the amino acid $\hat{\vect{s}}_q(l)$ at the $l$-th position using the probability $p\left(\hat{\vect{s}}_q(l) | \mathcal{S}_q\right)$
    \STATE {\bfseries Return:} Sequence $\hat{\vect{s}}_q$
\end{algorithmic}
\end{algorithm}

%% file: section/tab/tab-2-param.tex
\begin{table}[t]
\small
\caption{Hyperparameters of {\igseek}.}
\vspace{-2mm}
\centering
\begin{tabular}{ccl}
\toprule 
Hyperparameter  & Value & Description \\ 
\midrule
\multicolumn{3}{c}{Input}\\
\midrule
noise\_ratio   & $0.15$             & Ratio of the input coordinates with added Gaussian noise. \\
noise\_scale   & $1$                & The standard deviation $\sigma$ in the Gaussian noise. \\
$\theta$       & $10$ \AA\             & The Euclidean distance threshold when constructing the graph $G$. \\
\midrule
\multicolumn{3}{c}{MEGNN}\\
\midrule
learning\_rate & $5 \times 10^{-3}$ & Learning rate of MEGNN. \\
weight\_decay  & $1 \times 10^{-4}$ & Weight decay factor of the optimizer. \\
hidden\_dim    & $256$              & Size of hidden feature dimension in MEGNN. \\
emb\_dim       & $128$              & Size of output embedding dimension in MEGNN. \\
n\_layer       & $4$                & Number of layers in MEGNN. \\
epoch          & $50$               & Number of the iterations during training\\
batch\_size    & $8$                & Number of batch size in MEGNN. \\
drop\_out      & $0.1$              & Number of dropout rate in MEGNN.\\
\midrule
\multicolumn{3}{c}{Retrieval}\\
\midrule
$k$            & $10$               & Number of nearest neighbor retrieved in the CDR vector database. \\
n\_sample      & $2$                & Number of generated samples for each query.\\
\bottomrule
\end{tabular}
\label{tab:appendix-param}
\vspace{-3mm}
\end{table}

%% file: section/tab/tab-1-dataset.tex
\begin{table}[t]
\small
\caption{Profile of Datasets}
\vspace{-2mm}
\centering
\label{tab:datasets}
\begin{tabular}{crrrrrr}
\toprule 
{SAbDab}  & \#CDR-H1& \#CDR-H2& \#CDR-H3& \#CDR-L1& \#CDR-L2& \#CDR-L3\\ 
\midrule
 \# Data (before-2024) & 4,464     & 4,466    & 4,463    & 3,693        & 3,696  & 3,897  \\ 
\# Query (2024) & 809     & 823    & 513    & 580        & 607  & 578  \\ 
\midrule
\midrule
{STCRDab}  & \#CDR-A1& \#CDR-A2& \#CDR-A3& \#CDR-B1& \#CDR-B2& \#CDR-B3\\ 
\midrule
 \# Data  & 680  & 680    & 680 & 741        & 741       & 741         \\
\# Query    & 138    & 140   & 120 & 158        & 154       & 138      \\ 
\bottomrule
\end{tabular}
\vspace{-4mm}
\end{table}

%% file: section/tab/tab-4-K.tex
\begin{table}[t]
\small
\caption{The Comparison of Average AAR with varying $K$ in SAbDab-2024.}
\vspace{-2mm}
\centering
\label{tab:param-k}
\begin{tabular}{cccccc}
\toprule 
$K$  & 5 & 10 &	20 & 50 & 100 \\
\midrule
CDR-L1 & 0.660 & 0.658 & 0.645 & 0.620 & 0.593 \\
\midrule
CDR-L2 & 0.580 & 0.580 & 0.573 & 0.573 & 0.550 \\
\midrule
CDR-L3 & 0.586 & 0.586 & 0.576 & 0.574 & 0.564 \\
\midrule
CDR-H1 & 0.560 & 0.561 & 0.560 & 0.553 & 0.537 \\
\midrule
CDR-H2 & 0.440 & 0.435 & 0.432 & 0.429 & 0.430 \\
\midrule
CDR-H3 & 0.473 & 0.464 & 0.455 & 0.447 & 0.441 \\
\bottomrule
\end{tabular}
\vspace{-4mm}
\end{table}

%% file: igseek.bbl
\begin{thebibliography}{59}
\providecommand{\natexlab}[1]{#1}
\providecommand{\url}[1]{\texttt{#1}}
\expandafter\ifx\csname urlstyle\endcsname\relax
  \providecommand{\doi}[1]{doi: #1}\else
  \providecommand{\doi}{doi: \begingroup \urlstyle{rm}\Url}\fi

\bibitem[Abraham(2020)]{abraham2020passive}
Jonathan Abraham.
\newblock Passive antibody therapy in covid-19.
\newblock \emph{Nature Reviews Immunology}, 20\penalty0 (7):\penalty0 401--403, 2020.

\bibitem[Adams \& Weiner(2005)Adams and Weiner]{adams2005cancer}
Gregory~P Adams and Louis~M Weiner.
\newblock Monoclonal antibody therapy of cancer.
\newblock \emph{Nature biotechnology}, 23\penalty0 (9):\penalty0 1147--1157, 2005.

\bibitem[Adolf-Bryfogle et~al.(2015)Adolf-Bryfogle, Xu, North, Lehmann, and Dunbrack~Jr]{adolf2015fccc}
Jared Adolf-Bryfogle, Qifang Xu, Benjamin North, Andreas Lehmann, and Roland~L Dunbrack~Jr.
\newblock Pyigclassify: a database of antibody cdr structural classifications.
\newblock \emph{Nucleic acids research}, 43\penalty0 (D1):\penalty0 D432--D438, 2015.

\bibitem[Adolf-Bryfogle et~al.(2018)Adolf-Bryfogle, Kalyuzhniy, Kubitz, Weitzner, Hu, Adachi, Schief, and Dunbrack~Jr]{adolf2018rosetta}
Jared Adolf-Bryfogle, Oleks Kalyuzhniy, Michael Kubitz, Brian~D Weitzner, Xiaozhen Hu, Yumiko Adachi, William~R Schief, and Roland~L Dunbrack~Jr.
\newblock Rosettaantibodydesign (rabd): A general framework for computational antibody design.
\newblock \emph{PLoS computational biology}, 14\penalty0 (4):\penalty0 1--38, 2018.

\bibitem[Baek et~al.(2021)Baek, DiMaio, Anishchenko, Dauparas, Ovchinnikov, Lee, Wang, Cong, Kinch, Schaeffer, Mill{\'a}n, Park, Adams, Glassman, DeGiovanni, Pereira, Rodrigues, Dijk, Ebrecht, Opperman, Sagmeister, Buhlheller, Pavkov-Keller, Rathinaswamy, Dalwadi, Yip, Burke, Garcia, Grishin, Adams, Read, and Baker]{baek2021rosetta}
Minkyung Baek, Frank DiMaio, Ivan Anishchenko, Justas Dauparas, Sergey Ovchinnikov, Gyu~Rie Lee, Jue Wang, Qian Cong, Lisa~N Kinch, R~Dustin Schaeffer, Claudia Mill{\'a}n, Hahnbeom Park, Carson Adams, Caleb~R Glassman, Andy DeGiovanni, Jose~H Pereira, Andria~V Rodrigues, Alberdina A~Van Dijk, Ana~C Ebrecht, Diederik~J Opperman, Theo Sagmeister, Christoph Buhlheller, Tea Pavkov-Keller, Manoj~K Rathinaswamy, Udit Dalwadi, Calvin~K Yip, John~E Burke, K~Christopher Garcia, Nick~V Grishin, Paul~D Adams, Randy~J Read, and David Baker.
\newblock Accurate prediction of protein structures and interactions using a three-track neural network.
\newblock \emph{Science}, 373\penalty0 (6557):\penalty0 871--876, 2021.

\bibitem[Batzner et~al.(2022)Batzner, Musaelian, Sun, Geiger, Mailoa, Kornbluth, Molinari, Smidt, and Kozinsky]{batzner2022nequip}
Simon Batzner, Albert Musaelian, Lixin Sun, Mario Geiger, Jonathan~P Mailoa, Mordechai Kornbluth, Nicola Molinari, Tess~E Smidt, and Boris Kozinsky.
\newblock E(3)-equivariant graph neural networks for data-efficient and accurate interatomic potentials.
\newblock \emph{Nature communications}, 13\penalty0 (1):\penalty0 2453, 2022.

\bibitem[Bennett et~al.(2024)Bennett, Watson, Ragotte, Borst, See, Weidle, Biswas, Shrock, Leung, Huang, Goreshnik, Ault, Carr, Singer, Criswell, Vafeados, Sanchez, Kim, Torres, Chan, and Baker]{bennett2024atomically}
Nathaniel~R Bennett, Joseph~L Watson, Robert~J Ragotte, Andrew~J Borst, D{\'e}jena{\'e}~L See, Connor Weidle, Riti Biswas, Ellen~L Shrock, Philip~JY Leung, Buwei Huang, Inna Goreshnik, Russell Ault, Kenneth~D Carr, Benedikt Singer, Cameron Criswell, Dionne Vafeados, Mariana~Garcia Sanchez, Ho~Min Kim, Susana~V{\'a}zquez Torres, Sidney Chan, and David~and Baker.
\newblock Atomically accurate de novo design of single-domain antibodies.
\newblock \emph{bioRxiv}, 2024.

\bibitem[Chao et~al.(2006)Chao, Lau, Hackel, Sazinsky, Lippow, and Wittrup]{chao2006yeast}
Ginger Chao, Wai~L Lau, Benjamin~J Hackel, Stephen~L Sazinsky, Shaun~M Lippow, and K~Dane Wittrup.
\newblock Isolating and engineering human antibodies using yeast surface display.
\newblock \emph{Nature protocols}, 1\penalty0 (2):\penalty0 755--768, 2006.

\bibitem[Chothia et~al.(1989)Chothia, Lesk, Tramontano, Levitf, Smith-GiII, Air, Sheriff, Padlan, Davies, Tulip, Colman, Spinelli, Alzari, and Poljak]{chothia1989conformations}
Cyrus Chothia, Arthur~M. Lesk, Anna Tramontano, Michael Levitf, Sandra~J. Smith-GiII, Gillian Air, Steven Sheriff, Eduardo~A. Padlan, David Davies, William~R. Tulip, Peter~M. Colman, Silvia Spinelli, Pedro~M. Alzari, and Roberto~J. Poljak.
\newblock Conformations of immunoglobulin hypervariable regions.
\newblock \emph{Nature}, 342\penalty0 (6252):\penalty0 877--883, 1989.

\bibitem[Cybenko(1989)]{cybenko1989mlp}
George~V. Cybenko.
\newblock Approximation by superpositions of a sigmoidal function.
\newblock \emph{Mathematics of Control, Signals and Systems}, 2\penalty0 (4):\penalty0 303--314, 1989.

\bibitem[Dauparas et~al.(2022)Dauparas, Anishchenko, Bennett, Bai, Ragotte, Milles, Wicky, Courbet, de~Haas, Bethel, Leung, Huddy, Pellock, Tischer, Chan, Koepnick, Nguyen, Kang, Sankaran, Bera, King, and Baker]{dauparas2022mpnn}
J.~Dauparas, I.~Anishchenko, N.~Bennett, H.~Bai, R.~J. Ragotte, L.~F. Milles, B.~I.~M. Wicky, A.~Courbet, R.~J. de~Haas, N.~Bethel, P.~J.~Y. Leung, T.~F. Huddy, S.~Pellock, D.~Tischer, F.~Chan, B.~Koepnick, H.~Nguyen, A.~Kang, B.~Sankaran, A.~K. Bera, N.~P. King, and D.~Baker.
\newblock Robust deep learning–based protein sequence design using proteinmpnn.
\newblock \emph{Science}, 378\penalty0 (6615):\penalty0 49--56, 2022.

\bibitem[Dreyer et~al.(2023)Dreyer, Cutting, Schneider, Kenlay, and Deane]{dreyer2023abmpnn}
Fr{\'e}d{\'e}ric~A Dreyer, Daniel Cutting, Constantin Schneider, Henry Kenlay, and Charlotte~M Deane.
\newblock Inverse folding for antibody sequence design using deep learning.
\newblock In \emph{ICML CompBio}, 2023.

\bibitem[Dunbar et~al.(2013)Dunbar, Krawczyk, Leem, Baker, Fuchs, Georges, Shi, and Deane]{dunbar2013sabdab}
James Dunbar, Konrad Krawczyk, Jinwoo Leem, Terry Baker, Angelika Fuchs, Guy Georges, Jiye Shi, and Charlotte~M. Deane.
\newblock Sabdab: the structural antibody database.
\newblock \emph{Nucleic Acids Res.}, 42\penalty0 (D1):\penalty0 D1140--D1146, 2013.

\bibitem[Dunleavy(2024)]{dunleavy2024keytruda}
Kevin Dunleavy.
\newblock Who's no. 1? with \$25b in sales, merck's keytruda looks to be the top-selling drug of 2023.
\newblock \emph{Fierce Pharma}, 2024.

\bibitem[Feldmann \& Maini(2003)Feldmann and Maini]{feldmann2003tnf}
Marc Feldmann and Ravinder~N Maini.
\newblock Tnf defined as a therapeutic target for rheumatoid arthritis and other autoimmune diseases.
\newblock \emph{Nature medicine}, 9\penalty0 (10):\penalty0 1245--1250, 2003.

\bibitem[Frey \& Dueck(2007)Frey and Dueck]{frey2007clustering}
Brendan~J Frey and Delbert Dueck.
\newblock Clustering by passing messages between data points.
\newblock \emph{science}, 315\penalty0 (5814):\penalty0 972--976, 2007.

\bibitem[Fuchs et~al.(2020)Fuchs, Worrall, Fischer, and Welling]{fuchs2020se3transformer}
Fabian Fuchs, Daniel Worrall, Volker Fischer, and Max Welling.
\newblock Se(3)-transformers: 3d roto-translation equivariant attention networks.
\newblock In \emph{NeurIPS}, pp.\  1970--1981, 2020.

\bibitem[Funahashi(1989)]{funahashi1989mlp}
Ken-Ichi Funahashi.
\newblock On the approximate realization of continuous mappings by neural networks.
\newblock \emph{Neural Networks}, 2\penalty0 (3):\penalty0 183--192, 1989.

\bibitem[Gao et~al.(2023{\natexlab{a}})Gao, Xiong, Gao, Jia, Pan, Bi, Dai, Sun, Guo, Wang, and Wang]{DBLP:journals/corr/abs-2312-10997}
Yunfan Gao, Yun Xiong, Xinyu Gao, Kangxiang Jia, Jinliu Pan, Yuxi Bi, Yi~Dai, Jiawei Sun, Qianyu Guo, Meng Wang, and Haofen Wang.
\newblock Retrieval-augmented generation for large language models: {A} survey.
\newblock \emph{CoRR}, abs/2312.10997, 2023{\natexlab{a}}.

\bibitem[Gao et~al.(2023{\natexlab{b}})Gao, Tan, Chac{\'o}n, and Li]{gao2023pifold}
Zhangyang Gao, Cheng Tan, Pablo Chac{\'o}n, and Stan~Z Li.
\newblock Pifold: Toward effective and efficient protein inverse folding.
\newblock In \emph{ICLR}, 2023{\natexlab{b}}.

\bibitem[Gruver et~al.(2023)Gruver, Stanton, Frey, Rudner, Hotzel, Lafrance-Vanasse, Rajpal, Cho, and Wilson]{gruver2023lambo}
Nate Gruver, Samuel Stanton, Nathan Frey, Tim G.~J. Rudner, Isidro Hotzel, Julien Lafrance-Vanasse, Arvind Rajpal, Kyunghyun Cho, and Andrew~G Wilson.
\newblock Protein design with guided discrete diffusion.
\newblock In \emph{NeurIPS}, volume~36, pp.\  12489--12517, 2023.

\bibitem[He et~al.(2016)He, Zhang, Ren, and Sun]{he2016resnet}
Kaiming He, Xiangyu Zhang, Shaoqing Ren, and Jian Sun.
\newblock Deep residual learning for image recognition.
\newblock In \emph{CVPR}, pp.\  770--778, 2016.

\bibitem[H{\o}ie et~al.(2024)H{\o}ie, Hummer, Olsen, Nielsen, and Deane]{hoie2024antifold}
Magnus H{\o}ie, Alissa Hummer, Tobias Olsen, Morten Nielsen, and Charlotte Deane.
\newblock Antifold: Improved antibody structure design using inverse folding.
\newblock \emph{arXiv}, 2024.
\newblock URL \url{https://arxiv.org/abs/2405.03370}.

\bibitem[Holm(2020)]{holm2020dali}
Liisa Holm.
\newblock Using dali for protein structure comparison.
\newblock \emph{Structural Bioinformatics: Methods and Protocols}, pp.\  29--42, 2020.

\bibitem[Hornik(1991)]{hornik1991mlp}
Kurt Hornik.
\newblock Approximation capabilities of multilayer feedforward networks.
\newblock \emph{Neural Networks}, 4\penalty0 (2):\penalty0 251--257, 1991.

\bibitem[Hsu et~al.(2022)Hsu, Verkuil, Liu, Lin, Hie, Sercu, Lerer, and Rives]{hsu2022esmif1}
Chloe Hsu, Robert Verkuil, Jason Liu, Zeming Lin, Brian Hie, Tom Sercu, Adam Lerer, and Alexander Rives.
\newblock Learning inverse folding from millions of predicted structures.
\newblock In \emph{ICML}, pp.\  8946--8970, 2022.

\bibitem[Huang et~al.(2022)Huang, Han, Rong, Xu, Sun, and Huang]{huang2022gmn}
Wenbing Huang, Jiaqi Han, Yu~Rong, Tingyang Xu, Fuchun Sun, and Junzhou Huang.
\newblock Equivariant graph mechanics networks with constraints.
\newblock In \emph{ICLR}, 2022.

\bibitem[Jin et~al.(2022)Jin, Wohlwend, Barzilay, and Jaakkola]{jin2022refinegnn}
Wengong Jin, Jeremy Wohlwend, Regina Barzilay, and Tommi Jaakkola.
\newblock Iterative refinement graph neural network for antibody sequence-structure co-design.
\newblock In \emph{ICLR}, 2022.

\bibitem[Jing et~al.(2021)Jing, Eismann, Suriana, Townshend, and Dror]{jing2020gvp}
Bowen Jing, Stephan Eismann, Patricia Suriana, Raphael John~Lamarre Townshend, and Ron Dror.
\newblock Learning from protein structure with geometric vector perceptrons.
\newblock In \emph{International Conference on Learning Representations}, 2021.

\bibitem[Jumper et~al.(2021)Jumper, Evans, Pritzel, Green, Figurnov, Ronneberger, Tunyasuvunakool, Bates, {\v{Z}}{\'\i}dek, Potapenko, Bridgland, Meyer, Kohl, Ballard, Cowie, Romera-Paredes, Nikolov, Jain, Adler, Back, Petersen, Reiman, Clancy, Zielinski, Steinegger, Pacholska, Berghammer, Bodenstein, Silver, Vinyals, Senior, Kavukcuoglu, Kohli, and Hassabis]{jumper2021alphafold}
John Jumper, Richard Evans, Alexander Pritzel, Tim Green, Michael Figurnov, Olaf Ronneberger, Kathryn Tunyasuvunakool, Russ Bates, Augustin {\v{Z}}{\'\i}dek, Anna Potapenko, Alex Bridgland, Clemens Meyer, Simon~AA Kohl, Andrew~J Ballard, Andrew Cowie, Bernardino Romera-Paredes, Stanislav Nikolov, Rishub Jain, Jonas Adler, Trevor Back, Stig Petersen, David Reiman, Ellen Clancy, Michal Zielinski, Martin Steinegger, Michalina Pacholska, Tamas Berghammer, Sebastian Bodenstein, David Silver, Oriol Vinyals, Andrew~W Senior, Koray Kavukcuoglu, Pushmeet Kohli, and Demis Hassabis.
\newblock Highly accurate protein structure prediction with alphafold.
\newblock \emph{nature}, 596\penalty0 (7873):\penalty0 583--589, 2021.

\bibitem[Kingma \& Ba(2015)Kingma and Ba]{kingma2015adam}
Diederik~P Kingma and Jimmy~Lei Ba.
\newblock Adam: A method for stochastic optimization.
\newblock In \emph{ICLR}, 2015.

\bibitem[Kong et~al.(2023{\natexlab{a}})Kong, Huang, and Liu]{kong2023dymean}
Xiangzhe Kong, Wenbing Huang, and Yang Liu.
\newblock End-to-end full-atom antibody design.
\newblock In \emph{ICML}, pp.\  17409--17429, 2023{\natexlab{a}}.

\bibitem[Kong et~al.(2023{\natexlab{b}})Kong, Huang, and Liu]{kong2023mean}
Xiangzhe Kong, Wenbing Huang, and Yang Liu.
\newblock Conditional antibody design as 3d equivariant graph translation.
\newblock In \emph{ICLR}, 2023{\natexlab{b}}.

\bibitem[Kovaltsuk et~al.(2018)Kovaltsuk, Leem, Kelm, Snowden, Deane, and Krawczyk]{kovaltsuk2018oas}
Aleksandr Kovaltsuk, Jinwoo Leem, Sebastian Kelm, James Snowden, Charlotte~M Deane, and Konrad Krawczyk.
\newblock Observed antibody space: a resource for data mining next-generation sequencing of antibody repertoires.
\newblock \emph{J. Immunol.}, 201\penalty0 (8):\penalty0 2502--2509, 2018.

\bibitem[Leem et~al.(2018)Leem, de~Oliveira, Krawczyk, and Deane]{leem2018stcrdab}
Jinwoo Leem, Saulo H~P de~Oliveira, Konrad Krawczyk, and Charlotte~M Deane.
\newblock Stcrdab: the structural t-cell receptor database.
\newblock \emph{Nucleic acids research}, 46\penalty0 (D1):\penalty0 D406--D412, 2018.

\bibitem[Lefranc et~al.(2003)Lefranc, Pommi{\'e}, Ruiz, Giudicelli, Foulquier, Truong, Thouvenin-Contet, and Lefranc]{lefranc2003imgt}
Marie-Paule Lefranc, Christelle Pommi{\'e}, Manuel Ruiz, V{\'e}ronique Giudicelli, Elodie Foulquier, Lisa Truong, Val{\'e}rie Thouvenin-Contet, and G{\'e}rard Lefranc.
\newblock Imgt unique numbering for immunoglobulin and t cell receptor variable domains and ig superfamily v-like domains.
\newblock \emph{Developmental \& Comparative Immunology}, 27\penalty0 (1):\penalty0 55--77, 2003.

\bibitem[Liao \& Smidt(2023)Liao and Smidt]{liao2023equiformer}
Yi-Lun Liao and Tess Smidt.
\newblock Equiformer: Equivariant graph attention transformer for 3d atomistic graphs.
\newblock In \emph{ICLR}, 2023.

\bibitem[Luo et~al.(2022)Luo, Su, Peng, Wang, Peng, and Ma]{luo2022diffab}
Shitong Luo, Yufeng Su, Xingang Peng, Sheng Wang, Jian Peng, and Jianzhu Ma.
\newblock Antigen-specific antibody design and optimization with diffusion-based generative models for protein structures.
\newblock \emph{NeurIPS}, 35:\penalty0 9754--9767, 2022.

\bibitem[MacCallum et~al.(1996)MacCallum, Martin, and Thornton]{maccallum1996antibody}
Robert~M MacCallum, Andrew~CR Martin, and Janet~M Thornton.
\newblock Antibody-antigen interactions: contact analysis and binding site topography.
\newblock \emph{Journal of molecular biology}, 262\penalty0 (5):\penalty0 732--745, 1996.

\bibitem[North et~al.(2011)North, Lehmann, and Dunbrack~Jr]{north2011fccc}
Benjamin North, Andreas Lehmann, and Roland~L Dunbrack~Jr.
\newblock A new clustering of antibody cdr loop conformations.
\newblock \emph{Journal of molecular biology}, 406\penalty0 (2):\penalty0 228--256, 2011.

\bibitem[Notin et~al.(2024)Notin, Rollins, Gal, Sander, and Marks]{notin2024protein}
Pascal Notin, Nathan Rollins, Yarin Gal, Chris Sander, and Debora Marks.
\newblock Machine learning for functional protein design.
\newblock \emph{Nature biotechnology}, 42\penalty0 (2):\penalty0 216--228, 2024.

\bibitem[Olsen et~al.(2022)Olsen, Boyles, and Deane]{olsen2022oas}
Tobias~H Olsen, Fergus Boyles, and Charlotte~M Deane.
\newblock Observed antibody space: A diverse database of cleaned, annotated, and translated unpaired and paired antibody sequences.
\newblock \emph{Protein Science}, 31\penalty0 (1):\penalty0 141--146, 2022.

\bibitem[Paszke et~al.(2019)Paszke, Gross, Massa, Lerer, Bradbury, Chanan, Killeen, Lin, Gimelshein, Antiga, Desmaison, K{\"{o}}pf, Yang, DeVito, Raison, Tejani, Chilamkurthy, Steiner, Fang, Bai, and Chintala]{paszke2019pytorch}
Adam Paszke, Sam Gross, Francisco Massa, Adam Lerer, James Bradbury, Gregory Chanan, Trevor Killeen, Zeming Lin, Natalia Gimelshein, Luca Antiga, Alban Desmaison, Andreas K{\"{o}}pf, Edward~Z. Yang, Zachary DeVito, Martin Raison, Alykhan Tejani, Sasank Chilamkurthy, Benoit Steiner, Lu~Fang, Junjie Bai, and Soumith Chintala.
\newblock Pytorch: An imperative style, high-performance deep learning library.
\newblock In \emph{NeurIPS}, pp.\  8024--8035, 2019.

\bibitem[Proch{\'a}zka et~al.(2024)Proch{\'a}zka, Slanin{\'a}kov{\'a}, Olha, Ro{\v{s}}inec, Gre{\v{s}}ov{\'a}, J{\'a}no{\v{s}}ov{\'a}, {\v{C}}ill{\'\i}k, Porubsk{\'a}, Svobodov{\'a}, Dohnal, and Antol]{prochazka2024alphafind}
David Proch{\'a}zka, Ter{\'e}zia Slanin{\'a}kov{\'a}, Jaroslav Olha, Adri{\'a}n Ro{\v{s}}inec, Katar{\'\i}na Gre{\v{s}}ov{\'a}, Miriama J{\'a}no{\v{s}}ov{\'a}, Jakub {\v{C}}ill{\'\i}k, Jana Porubsk{\'a}, Radka Svobodov{\'a}, Vlastislav Dohnal, and Matej Antol.
\newblock Alphafind: discover structure similarity across the proteome in alphafold db.
\newblock \emph{Nucleic Acids Research}, 52\penalty0 (W1):\penalty0 W182--W186, 2024.

\bibitem[Satorras et~al.(2021)Satorras, Hoogeboom, and Welling]{satorras2021egnn}
V{\i}ctor~Garcia Satorras, Emiel Hoogeboom, and Max Welling.
\newblock E(n) equivariant graph neural networks.
\newblock In \emph{ICML}, pp.\  9323--9332, 2021.

\bibitem[Schneider et~al.(2021)Schneider, Raybould, and Deane]{schneider2021sabdab}
Constantin Schneider, Matthew I~J Raybould, and Charlotte~M Deane.
\newblock Sabdab in the age of biotherapeutics: updates including sabdab-nano, the nanobody structure tracker.
\newblock \emph{Nucleic Acids Res.}, 50\penalty0 (D1):\penalty0 D1368--D1372, 2021.

\bibitem[Shindyalov \& Bourne(1998)Shindyalov and Bourne]{shindyalov1998ce}
Ilya~N Shindyalov and Philip~E Bourne.
\newblock Protein structure alignment by incremental combinatorial extension (ce) of the optimal path.
\newblock \emph{Protein engineering}, 11\penalty0 (9):\penalty0 739--747, 1998.

\bibitem[Srivastava et~al.(2014)Srivastava, Hinton, Krizhevsky, Sutskever, and Salakhutdinov]{srivastava2014dropout}
Nitish Srivastava, Geoffrey Hinton, Alex Krizhevsky, Ilya Sutskever, and Ruslan Salakhutdinov.
\newblock Dropout: a simple way to prevent neural networks from overfitting.
\newblock \emph{JMLR}, 15\penalty0 (1):\penalty0 1929--1958, 2014.

\bibitem[van~den Oord et~al.(2017)van~den Oord, Vinyals, and Kavukcuoglu]{oord2017vqvae}
A{\"{a}}ron van~den Oord, Oriol Vinyals, and Koray Kavukcuoglu.
\newblock Neural discrete representation learning.
\newblock In \emph{NeurIPS}, pp.\  6306--6315, 2017.

\bibitem[Van~der Maaten \& Hinton(2008)Van~der Maaten and Hinton]{van2008tsne}
Laurens Van~der Maaten and Geoffrey Hinton.
\newblock Visualizing data using t-sne.
\newblock \emph{JMLR}, 9\penalty0 (11), 2008.

\bibitem[Van~Kempen et~al.(2024)Van~Kempen, Kim, Tumescheit, Mirdita, Lee, Gilchrist, S{\"o}ding, and Steinegger]{van2024foldseek}
Michel Van~Kempen, Stephanie~S Kim, Charlotte Tumescheit, Milot Mirdita, Jeongjae Lee, Cameron~LM Gilchrist, Johannes S{\"o}ding, and Martin Steinegger.
\newblock Fast and accurate protein structure search with foldseek.
\newblock \emph{Nature biotechnology}, 42\penalty0 (2):\penalty0 243--246, 2024.

\bibitem[Van~Wauwe et~al.(1980)Van~Wauwe, De~Mey, and Goossens]{van1980okt3}
Jean~P Van~Wauwe, JR~De~Mey, and JG~Goossens.
\newblock Okt3: a monoclonal anti-human t lymphocyte antibody with potent mitogenic properties.
\newblock \emph{Journal of immunology (Baltimore, Md.: 1950)}, 124\penalty0 (6):\penalty0 2708--2713, 1980.

\bibitem[Varadi et~al.(2022)Varadi, Anyango, Deshpande, Nair, Natassia, Yordanova, Yuan, Stroe, Wood, Laydon, {\v{Z}}{\'\i}dek, Green, Tunyasuvunakool, Petersen, Jumper, Clancy, Green, Vora, Lutfi, Figurnov, Cowie, Hobbs, Kohli, Kleywegt, Birney, Hassabis, and Velankar]{varadi2022alphafolddb}
Mihaly Varadi, Stephen Anyango, Mandar Deshpande, Sreenath Nair, Cindy Natassia, Galabina Yordanova, David Yuan, Oana Stroe, Gemma Wood, Agata Laydon, Augustin {\v{Z}}{\'\i}dek, Tim Green, Kathryn Tunyasuvunakool, Stig Petersen, John Jumper, Ellen Clancy, Richard Green, Ankur Vora, Mira Lutfi, Michael Figurnov, Andrew Cowie, Nicole Hobbs, Pushmeet Kohli, Gerard Kleywegt, Ewan Birney, Demis Hassabis, and Sameer Velankar.
\newblock Alphafold protein structure database: massively expanding the structural coverage of protein-sequence space with high-accuracy models.
\newblock \emph{Nucleic acids research}, 50\penalty0 (D1):\penalty0 D439--D444, 2022.

\bibitem[Wang et~al.(2024)Wang, Wu, Gao, Wu, Zhao, and Yao]{wang2024iggm}
Rubo Wang, Fandi Wu, Xingyu Gao, Jiaxiang Wu, Peilin Zhao, and Jianhua Yao.
\newblock Iggm: A generative model for functional antibody and nanobody design.
\newblock \emph{bioRxiv}, 2024.

\bibitem[Watson et~al.(2023)Watson, Juergens, Bennett, Trippe, Yim, Eisenach, Ahern, Borst, Ragotte, Milles, Wicky, Hanikel, Pellock, Courbet, Sheffler, Wang, Venkatesh, Sappington, Torres, Lauko, De~Bortoli, Mathieu, Ovchinnikov, Barzilay, Jaakkola, DiMaio, Baek, and Baker]{watson2023rfdiffusion}
Joseph~L Watson, David Juergens, Nathaniel~R Bennett, Brian~L Trippe, Jason Yim, Helen~E Eisenach, Woody Ahern, Andrew~J Borst, Robert~J Ragotte, Lukas~F Milles, Basile I~M Wicky, Nikita Hanikel, Samuel~J Pellock, Alexis Courbet, William Sheffler, Jue Wang, Preetham Venkatesh, Isaac Sappington, Susana~V{\'a}zquez Torres, Anna Lauko, Valentin De~Bortoli, Emile Mathieu, Sergey Ovchinnikov, Regina Barzilay, Tommi~S Jaakkola, Frank DiMaio, Minkyung Baek, and David Baker.
\newblock De novo design of protein structure and function with rfdiffusion.
\newblock \emph{Nature}, 620\penalty0 (7976):\penalty0 1089--1100, 2023.

\bibitem[Xu et~al.(2015)Xu, Wang, Chen, and Li]{xu2015leakyrelu}
Bing Xu, Naiyan Wang, Tianqi Chen, and Mu~Li.
\newblock Empirical evaluation of rectified activations in convolutional network.
\newblock \emph{arXiv}, 2015.
\newblock URL \url{https://arxiv.org/abs/1505.00853}.

\bibitem[Zhang et~al.(2022)Zhang, Shine, Pyle, and Zhang]{zhang2022usalign}
Chengxin Zhang, Morgan Shine, Anna~Marie Pyle, and Yang Zhang.
\newblock Us-align: universal structure alignments of proteins, nucleic acids, and macromolecular complexes.
\newblock \emph{Nature methods}, 19\penalty0 (9):\penalty0 1109--1115, 2022.

\bibitem[Zhang \& Skolnick(2005)Zhang and Skolnick]{zhang2005tmalign}
Yang Zhang and Jeffrey Skolnick.
\newblock Tm-align: a protein structure alignment algorithm based on the tm-score.
\newblock \emph{Nucleic acids research}, 33\penalty0 (7):\penalty0 2302--2309, 2005.

\bibitem[Zhang et~al.(2023)Zhang, Xu, Jamasb, Chenthamarakshan, Lozano, Das, and Tang]{zhang2023gearnet}
Zuobai Zhang, Minghao Xu, Arian Jamasb, Vijil Chenthamarakshan, Aurelie Lozano, Payel Das, and Jian Tang.
\newblock Protein representation learning by geometric structure pretraining.
\newblock In \emph{ICLR}, 2023.

\end{thebibliography}
